\documentclass[prd,nofootinbib,superscriptaddress,floatfix]{revtex4-1}

\makeatletter
\DeclareRobustCommand*{\bfseries}{%
  \not@math@alphabet\bfseries\mathbf
  \fontseries\bfdefault\selectfont
  \boldmath
}
\makeatother

\usepackage{calc}
\usepackage{rotating}
\usepackage[english]{babel}
\usepackage[T1]{fontenc}
\usepackage{textcomp}
\usepackage[osf,slantedGreek]{mathpazo}
\usepackage{graphicx}
\usepackage{subfig}
\usepackage{float}
\usepackage{amsmath}
\usepackage{amssymb}
\usepackage{amsthm}
\usepackage{latexsym}
\usepackage{dcolumn}

\newcommand*{\eg}{\textrm{e.g.}} 

\newcommand*{\etc}{\textrm{etc}}

\newcommand*{\alphaemmz}{\ensuremath{1/\alpha_{\text{em}}(M_Z)^{\overline{MS}}}}
\newcommand*{\alphas}{\ensuremath{\alpha_s(M_Z)^{\overline{MS}}}}
\newcommand*{\mb}{\ensuremath{m_b (m_b)^{\overline{MS}}}}
\newcommand*{\mt}{\ensuremath{m_t^{pole}}} 

\newcommand*{\mhalf}{\ensuremath{m_{1/2}}}
\newcommand*{\mzero}{\ensuremath{m_0}}
\newcommand*{\tanb}{\ensuremath{\tan\beta}}
\newcommand*{\azero}{\ensuremath{A_0}}
\newcommand*{\signmu}{\ensuremath{\text{sgn}\,\mu}}

\newcommand*{\relic}{\ensuremath{\Omega h^2}}
\newcommand*{\sineff}{\ensuremath{\sin^2 {\theta}_{\rm eff}}}
\newcommand*{\mw}{\ensuremath{M_W}}
\newcommand*{\deltaa}{\ensuremath{\delta(g-2)_{\mu}^{\rm SUSY}}}
\newcommand*{\brbsmumu}{\ensuremath{\mathcal{BR}\left(B_s \rightarrow \mu^+ \mu^-\right)}}
\newcommand*{\brbsg}{\ensuremath{\mathcal{BR}\left(\bar{B} \rightarrow X_s \gamma \right)}}
\newcommand*{\brbtn}{\ensuremath{\mathcal{BR}\left(B_u \rightarrow \tau \nu\right)}}
\newcommand*{\bsmumu}{\ensuremath{B_s \rightarrow \mu^+ \mu^-}}
\newcommand*{\bsg}{\ensuremath{\bar{B} \rightarrow X_s \gamma}}
\newcommand*{\btn}{\ensuremath{B_u \rightarrow \tau \nu}}
\newcommand*{\deltam}{\ensuremath{\Delta M_{B_s}}}
\newcommand*{\sigsip}{\ensuremath{\sigma_p^{\rm SI}}}
\newcommand*{\mh}{\ensuremath{m_h}}

\newcommand*{\neutone}{\ensuremath{\chi}}

\newcommand*{\charone}{\ensuremath{{\chi}^{\pm}_1}}

\newcommand*{\ETslash}{\ensuremath{/ \hspace{-.7em} E_T}}

\newcommand\lsim{\mathrel{\rlap{\lower4pt\hbox{\hskip1pt$\sim$}}
   \raise1pt\hbox{$<$}}}
\newcommand\gsim{\mathrel{\rlap{\lower4pt\hbox{\hskip1pt$\sim$}}
   \raise1pt\hbox{$>$}}}

\newcommand*{\neutralinoone}{\neutone}

\newcommand*{\charginoone}{\charone}

\newcommand*{\squark}{\ensuremath{\tilde{q}}}

\newcommand*{\gluino}{\ensuremath{\tilde{g}}} 
\newcommand*{\mgluino}{\ensuremath{{m_{\gluino}}}}

\newcommand*{\GeV}{\ensuremath{\,\text{GeV}}}
\newcommand*{\TeV}{\ensuremath{\,\text{TeV}}}
\newcommand*{\invpb}{\ensuremath{/\text{pb}}}
\newcommand*{\invfb}{\ensuremath{/\text{fb}}}

\newcommand*{\alphaT}{\ensuremath{\alpha_T}}

\newcommand*{\alphaTexp}{\ensuremath{\cms\ \alphaT\ 1.1\invfb} analysis}
\newcommand*{\alphaTonefb}{\ensuremath{\cms\ \alphaT\ 1.1\invfb} }
\newcommand*{\alphaTlim}{\ensuremath{\cms\ \alphaT\ 1.1\invfb} limit}

\newcommand*{\bayesfits}{\textrm{BayesFits Group}}
\newcommand*{\softsusy}{SoftSUSY}
\newcommand*{\superbayes}{\text{SuperBayeS}}

\newcommand*{\susyhit}{\text{SUSY-HIT}}
\newcommand*{\atlas}{\text{ATLAS}}
\newcommand*{\cms}{\text{CMS}}

\newcommand*{\pythia}{\text{PYTHIA}}
\newcommand*{\xenon}{\text{XENON100}}

\newcommand*{\Ht}{\ensuremath{H_T}}
\newcommand*{\et}{\ensuremath{E_T}}
\newcommand*{\pt}{\ensuremath{p_T}}

\let\oldcite\cite
\renewcommand*{\cite}{~\oldcite}
\newcommand*{\reftable}[1]{Table~\ref{#1}}
\newcommand*{\reffigure}[1]{Fig.~\ref{#1}}
\newcommand*{\refequation}[1]{Eq.~\ref{#1}}
\newcommand*{\refsection}[1]{Sec.~\ref{#1}}
\newcommand*{\refref}[1]{Ref.\cite{#1}}
\newcommand*{\refrefs}[1]{Refs.\cite{#1}}
\newcommand{\datalr}{d}

\newcommand*{\hl}{\ensuremath{h}}

\newcommand*{\ha}{\ensuremath{A}}

\newcommand*{\pb}{\ensuremath{\, \text{pb}}}
\newcommand*{\fb}{\ensuremath{\, \text{fb}}}

\newcommand*{\abundchi}{\ensuremath{\Omega_\chi h^2}}

\restylefloat{figure}
\newcommand*{\scale}{1}

\begin{document}

\title{Bayesian Implications of Current LHC and XENON100 Search Limits
  for the Constrained MSSM}

\author{Andrew Fowlie}
\email{A.Fowlie@sheffield.ac.uk}
\affiliation{Department of Physics and Astronomy, University of
  Sheffield, Sheffield S3 7RH, United Kingdom}

\author{Artur Kalinowski}
\email{Artur.Kalinowski@fuw.edu.pl}
\affiliation{Department of Physics, University of Warsaw, Ho{\. z}a 69, 00-681 Warsaw, Poland}

\author{Malgorzata Kazana}
\email{Malgorzata.Kazana@fuw.edu.pl}
\affiliation{National Centre for Nuclear Research, Ho{\. z}a
  69, 00-681 Warsaw, Poland}

\author{Leszek Roszkowski\footnote{On leave of absence from Department
    of Physics and Astronomy, University of Sheffield.}  }
\email{L.Roszkowski@sheffield.ac.uk} \affiliation{National Centre for
  Nuclear Research, Ho{\. z}a 69, 00-681 Warsaw, Poland}
\affiliation{Department of Physics and Astronomy, University of
  Sheffield, Sheffield S3 7RH, England}

\author{Y.-L. Sming Tsai}
\email{Sming.Tsai@fuw.edu.pl}
\affiliation{National Centre for Nuclear Research, Ho{\. z}a 69, 00-681 Warsaw, Poland}

\collaboration{\bayesfits}

\date{\today}

\begin{abstract}
  The \cms\ Collaboration has released the results of its search for
  supersymmetry, by applying an \alphaT\ method to $1.1\invfb$ of data
  at $7\TeV$.  The null result excludes (at $95$\%~C.L.)
  a low-mass region of the Constrained MSSM's parameter space that was previously 
  favored by other experiments.  Additionally, the negative result of
  the \xenon\ dark matter search has excluded (at $90$\%~C.L.) values of the
  spin-independent scattering cross sections \sigsip\ as low as
  $10^{-8}\pb$. We incorporate these improved experimental constraints into
  a global Bayesian fit of the Constrained MSSM by constructing
  approximate likelihood functions. In the case of the \alphaT\ limit,
  we simulate detector efficiency for the \alphaTexp\ and validate our
  method against the official $95\%$~C.L. contour. We identify the
  $68$\% and $95$\% credible posterior regions of the CMSSM parameters, and also
  find the best-fit point. 
  We find that the credible regions change considerably once a
  likelihood from \alphaT\ is included, in particular the narrow light
  Higgs resonance region becomes excluded, but the focus point/horizontal branch
  region remains allowed at the $1\sigma$ level. Adding the limit from
  \xenon\ has a weaker additional effect, in part due to large
  uncertainties in evaluating \sigsip, which we include in a
  conservative way, although we find that it reduces the posterior
  probability of the focus point region to the $2\sigma$ level. The
  new regions of high posterior favor squarks lighter than the gluino
  and all but one Higgs bosons heavy. The dark matter neutralino mass
  is found in the range $250\GeV\lsim m_\neutone \lsim 343\GeV$ (at
  $1\sigma$) while, as the result of improved limits from the LHC,
  the favored range of \sigsip\ is pushed down to values below
  $10^{-9}\pb$.  We highlight tension between \deltaa\ and \brbsg,
  which is exacerbated by including the \alphaT\ limit; each
  constraint favors a different region of the CMSSM's mass
  parameters.

\end{abstract}
\maketitle

\section{\label{sec:introduction}Introduction}
The search for new physics at the Large Hadron Collider (LHC) began in
earnest last year. An initial dataset of about $35\invpb$ at $\sqrt{s}
= 7\TeV$ was followed this year by a much larger collection of
about $5\invfb$ of data.  Based on about $1\invfb$ of analyzed
data, earlier this year new limits were published by
\atlas\cite{Aad:2011ib} and \cms\cite{Chatrchyan:2011zy} experimental
collaborations which significantly improved early LEP\cite{lepsusy}
and recent Tevatron results\cite{Abazov:2009zi,*Aaltonen:2008pv,*Aaltonen:2010uf}
as well as their own initial exclusion limits on the mass scale of
low-energy supersymmetry (SUSY), in particular, on the mass parameter
space of the Constrained MSSM
(CMSSM)\cite{Kane:1993td,*Martin:1997ns}.

The CMSSM is a tractable unified model of effective softly broken
supersymmetry, which includes models like, \eg, a relaxed version of
the minimal supergravity model\cite{Chamseddine:1982jx,*Nath:1983aw,*Hall:1983iz}. Despite its
restrictive boundary conditions, the CMSSM has a rich phenomenology
and has been extensively used as a basis for evaluating prospects for
SUSY searches at the LHC and in other collider, noncollider and dark
matter (DM) experiments. The model is defined by four continuous
parameters and one sign\cite{Kane:1993td}: \mzero, the universal
scalar mass; \mhalf, the universal gaugino mass; \azero, the universal
trilinear coupling; \tanb, the ratio of the Higgs vacuum expectation
values; and \signmu; the sign of the Higgs/Higgsino mass parameter
$\mu$. We denote them collectively by $\theta$.

So far, the best limits from the LHC experiments \atlas\ and \cms\
have come from channels involving only hadronic final states, with
\atlas\ investigating jets plus missing transverse momentum, and \cms\
using jets plus missing transverse energy, a ``razor'' and
\alphaT\cite{Lungu:2009nh,Chatrchyan:2011zy} search methods. In particular, they
involve different ways of clustering events into effective dijet
systems. Based on $1.1\invfb$ of data the strongest bounds on the
scale of SUSY have so far come from the \alphaT\ method.

Even before the turn-on of the LHC, the parameter space of the CMSSM
had been significantly
constrained\cite{Allanach:2005kz,*Allanach:2006jc,deAustri:2006pe,Allanach:2007qk,Feroz:2008wr,Buchmueller:2010ai,Trotta:2006ew,*Allanach:2006cc,*Buchmueller:2007zk,*Buchmueller:2008qe,Trotta:2008bp,*Feroz:2009dv,*Roszkowski:2009ye,*Buchmueller:2009fn}
by a variety of experimental data, most notably by LEP bounds on
electroweak observables, masses of the Standard Model (SM) and SM like
Higgs boson \hl\ and the lighter chargino \charginoone\cite{lepsusy},
data on heavy flavour processes: \bsg, \btn\ and \bsmumu,  the
difference between experimental and Standard Model contributions to the
anomalous magnetic moment of the muon
\deltaa\cite{Nakamura:2010zzi,*Miller:2007kk}, as well as the relic
abundance \abundchi\cite{Komatsu:2010fb} of the lightest neutralino,
which is assumed to be the dominant component of cold dark matter in the Universe.

In particular, a $\chi^2$ method favored a region of the CMSSM's
parameter space with $\mhalf \sim$ a few hundred \GeV\ and $\mzero
\alt \mhalf$\cite{Buchmueller:2010ai}. In a Bayesian approach, this
region was also favored but in addition a region of parameter space
with substantial posterior probability was also found in the so-called
horizontal branch, or focus point\cite{Chan:1997bi,*Feng:1999zg}, region of large $\mzero \agt
1\TeV$ and $\mhalf \ll \mzero$ and in, partly overlapping with it,
light Higgs resonance region\cite{deAustri:2006pe,Trotta:2008bp}.
A region of sizable posterior probability also exists in between
those two ``islands'' of the highest probability.

Whilst the first results from the LHC had a fairly mild effect on both
regions\cite{Buchmueller:2011aa,*Allanach:2011wi,Allanach:2011ut,Bertone:2011nj},
the current limits based on the early 2011 dataset of $\sim 1/\fb$
published so far take a much deeper bite into the (\mzero, \mhalf)
plane of the
CMSSM\cite{Buchmueller:2011sw,*Farina:2011bh,*Profumo:2011zj,Bertone:2011nj}.

Searches for signals of DM in direct detection (DD) experiments have
over the last few years also led to much improved limits
\cite{Angle:2009xb,*Ahmed:2009zw}.  Most notably, last Spring,
\xenon\cite{Aprile:2011dd,Aprile:2011hi}, with $100.9$ days of data,
excluded (at $90\%$~C.L.) spin-independent (SI) scattering cross sections
\sigsip\ as low as $10^{-8}\pb$, for some dark matter particle
masses\cite{Aprile:2011dd}.  The impact on the CMSSM's parameter space
of this new \xenon\ limit has been investigated in combination with
2010 LHC limits in
\refrefs{Bertone:2011nj,Buchmueller:2011ki,Akula:2011dd} and in
combination with current LHC limits in \refref{Buchmueller:2011sw}.
Moreover, LHCb has recently reported an upper bound on \brbsmumu\cite{Aaij:2011rja}
that is smaller than the previous best upper bound. 

In this paper, our goal is to perform a Bayesian analysis of the
CMSSM's parameter space that  includes the currently strongest LHC
limits on the SUSY mass scale and the \xenon\ limit on the DM scattering
cross section by carefully estimating associated uncertainties and including
them in the likelihood function. 
Combining limits from different search channels at the
LHC is a rather challenging task. In our approach we focus on the
result derived by \cms\ from the \alphaT\ method applied to $1.1\invfb$
of data, since it currently provides the strongest constraint on the
(\mzero, \mhalf) plane of the CMSSM
\cite{Buchmueller:2011ki,Akula:2011dd,Bechtle:2011it,Bertone:2011nj}.
We will simulate the \alphaT\ experiment at the event level, assume
that the experiment was a Poisson process and construct a likelihood
map on the CMSSM's (\mzero, \mhalf) plane.  Our approach is similar to
that of \refref{Allanach:2011ut}; however, it differs from that of
\refref{Buchmueller:2011aa,Buchmueller:2011sw,Buchmueller:2011ki}, in
which the likelihood is modeled with an empirical formula, and from
that of \refref{Bertone:2011nj}, in which the likelihood for the CMS
limit is approximated with a step function. In the latter approaches,
the likelihood is constructed from the published \cms\ \alphaT\ $95\%$
contour. In contrast, in our approach, the likelihood is constructed
in the whole (\mzero, \mhalf) plane
and next validated against the official \cms\ \alphaT\ $95\%$ contour.

This paper is organized as follows. In \refsection{sec:method} we
detail our methodology, including our statistical tools, scanning
algorithm, and our treatment of the likelihood from the \alphaTexp. In
\refsection{sec:results} we present results of scans that include
likelihoods from the \alphaTexp\ and from \xenon. We summarise our findings in
\refsection{sec:summary}.

\section{\label{sec:method}Method}
\subsection{\label{subsec:statlang}The statistical framework}
Our goal is to identify the regions of the CMSSM's parameter space
that are in best agreement with all relevant experimental constraints,
including the \alphaT\ and \xenon\ limits. The mass spectra and
other observables are also dependent on Standard Model parameters,
most notably the top pole mass \mt, the bottom mass \mb, the strong
coupling \alphas, and \alphaemmz. These ``nuisance'' parameters, which
we collectively denote by $\phi$, have been shown to play a
significant role in a statistical treatment\cite{Allanach:2004ed,deAustri:2006pe,Allanach:2007qk,Trotta:2008bp,Allanach:2005kz,*Allanach:2006jc}.

To set the stage, we define the best-fitting regions with Bayesian and
frequentist statistics. In both approaches
one considers the likelihood -- the probability of
obtaining experimental data for observables given the CMSSM's underlying parameters,
\begin{equation}
\mathcal{L}(\theta, \phi) = p(\datalr|\theta, \phi).
\end{equation} 
To see the likelihood's dependence on a particular parameter or set of
parameters, one maximizes over the CMSSM's other parameters, to obtain the
profile likelihood,
\begin{equation}
\mathcal{L}(\theta) = \max_{ \phi} p(\datalr|\theta, \phi).
\end{equation} 

In Bayesian statistics one addresses the question of the posterior 
-- what is the probability of the CMSSM's parameter values
given the experimental data? 
To this end one employs  Bayes's theorem to
find the posterior probability density function (pdf);
\begin{equation}\label{eqn:Bayes_theorem}
p(\theta,\phi|\datalr)=
\frac{{\mathcal L}(\theta,\phi) \pi(\theta,\phi)}{p(\datalr)}.
\end{equation} 
The Bayesian approach requires that we articulate our prior knowledge of the CMSSM's
parameters in the prior, $\pi(\theta,\phi)$. 
Finally, the denominator, $p(\datalr)$ is the evidence, which, because we are
not interested in model comparison, is merely a normalization factor.

To see the posterior's dependence on a particular parameter, or set of
parameters, one integrates, or marginalizes, over the CMSSM's other
parameters, as well as SM nuisance parameters, to obtain the
marginalized posterior pdf.

The two-dimensional region of the CMSSM's,  \eg, parameter space
(\mzero, \mhalf) that is in best agreement with the experiments, with
respect to the posterior -- the credible region -- is the smallest
region, $R$, that contains a given fraction of the posterior, that is,
the smallest region such that
\begin{equation}\label{eqn:cred_region_2D}
\int \limits_R \! p(\mzero, \mhalf | \datalr)  \, \text{d}\mzero \, \text{d}\mhalf = 1 - \epsilon.
\end{equation} 

The one-dimensional credible region, however, is the region such that the
posterior probability of the parameter being above the region is equal to the
posterior probability of the parameter being below the region and equal to a
given fraction of the posterior. That is, the one-dimensional credible region
from $L$ to $U$ of, \eg, \mzero, satisfies
\begin{equation}\label{eqn:cred_region_1D}
\int \limits_0^L \! p(\mzero|\datalr)  \, \text{d}\mzero = \int
\limits_U^\infty \! p(\mzero|\datalr)  \, \text{d}\mzero = \frac{1}{2} \epsilon.
\end{equation} 

In the frequentist approach, the $k$-dimensional region of the CMSSM's
parameter space that is in best agreement with the experiments, with
respect to the likelihood -- the confidence interval -- is the
region in which the $\chi^2$ is within $\Delta \chi^2$ of the minimum
$\chi^2$, where $\Delta \chi^2$ is such that
\begin{equation}\label{eqn:conf_interval}
F\left(\Delta \chi^2, k \right) = 1 - \epsilon,
\end{equation}
and $F\left(\Delta \chi^2, k \right)$ is a cumulative
$\chi^2$ distribution with $k$ degrees of freedom.

The parameter point with the minimum $\chi^2$ (or, equivalently, with
the maximum likelihood) is the best-fit point. In frequentist
statistics, the best-fit point has a special significance; the
confidence intervals are constructed from the best-fit point. In
contrast, the best-fit point has no significance in Bayesian
statistics.

From the best-fit point, one can obtain a $p$-value: the probability
of obtaining a $\chi^2$ value from experimental measurements equal or
larger than the best-fit $\chi^2$, accounting for the number of
degrees of freedom in the fit,
\begin{equation}\label{eqn:pvalue}
p\text{-value} = 1 - F\left(\chi^2, n \right),
\end{equation}
where $n$, the number of degrees of freedom in the fit, is the number
of experimental constraints in the $\chi^2$ calculation minus the
number of model parameters that were fitted.

Our credible regions and confidence intervals are defined by
\refequation{eqn:cred_region_2D}, \refequation{eqn:cred_region_1D} and
\refequation{eqn:conf_interval}, with $\epsilon 
= 0.32$ for $1\sigma$ and $\epsilon = 0.05$ for $2\sigma$. In
\refequation{eqn:conf_interval}, for a two-dimensional confidence
interval, this corresponds to $\Delta \chi^2 = 2.30$ for $1\sigma$ and
$\Delta \chi^2 = 5.99$ for $2\sigma$. 

Scanning the CMSSM's parameter space is computationally intensive --
a simple grid-scan of an eight-dimensional space is impractical. To
scan the CMSSM's parameter space efficiently, we use a modern (2006)
Monte Carlo algorithm, called Nested Sampling\cite{Feroz:2008xx},
which is tailored to work with Bayesian statistics.  We chose the
Nested Sampling settings so that the algorithm would accurately map
the posterior, but not necessarily the likelihood, in a reasonable CPU time
($4000$ live points and a stopping condition of $0.5$).

\begin{table}
\begin{tabular}{|l|l|l|l|}
\hline 
Parameter & Description & Prior Range & Scale \\
\hline 
\multicolumn{4}{|l|}{CMSSM} \\
\hline 
\mzero        	& Universal scalar mass          & $100$, $2000$ 	& Log\\
\mhalf		& Universal gaugino mass         & $100$, $1000$ 	& Log\\
\azero        	& Universal trilinear coupling   & $-2000$, $2000$	& Linear\\
\tanb	        & Ratio of Higgs VEVs            & $3$, $62$ 		& Linear\\
\signmu		& Sign of Higgs parameter        & $+1$ 		& Fixed\\
\hline 
\multicolumn{4}{|l|}{Nuisance} \\
\hline 
\mt           	& Top quark pole mass 	& $163.7$, $178.1$ 	& Linear\\
\mb 		& Bottom quark mass	& $3.92$, $4.48$ 	& Linear\\
\alphas	 	& Strong coupling	& $0.1096$, $0.1256$   	& Linear\\
\alphaemmz 	& Reciprocal of electromagnetic coupling  	& $127.846$, $127.99$ 	& Linear\\
\hline 
\end{tabular}
\caption{Priors for the CMSSM's parameters and for the Standard
  Model's nuisance parameters that we used in our scans. Masses are in
  GeV.}
\label{tab:priors}
\end{table}

Because of our lack of prior knowledge of the CMSSM's parameters, we
invoke the principle of insufficient reason and, as previously
in\cite{deAustri:2006pe,Roszkowski:2006mi,*Roszkowski:2007fd,Trotta:2008bp,LopezFogliani:2009np}
we choose noninformative priors for the CMSSM's parameters that
equally weight either linear or logarithmic intervals.  The priors
that we choose for \mzero\ and \mhalf\ equally weight logarithmic
intervals (log priors). This choice has been shown\cite{Trotta:2008bp}
not to suffer from the volume effect, unlike the flat prior, and it
also reduces the amount of fine tuning needed to achieve radiative
electroweak symmetry breaking.  For \tanb, \azero\ and for the SM's
nuisance parameters we choose equally weighted linear intervals
(linear priors). The prior ranges of the CMSSM's parameters and of the
SM's nuisance parameters over which we scan are listed in
\reftable{tab:priors}.

We use our updated and modified version of the \superbayes\cite{deAustri:2006pe,Trotta:2008bp} 
computer program to perform four scans of the CMSSM:
\begin{enumerate}
\item To set the stage for examining the impact of LHC and \xenon\
  limits, a scan involving constraints from only non-LHC experiments,
  including those listed above (see \reftable{tab:exp_constraints} for
  a complete list of observables and their values) but without a
  likelihood from \xenon.

\item To validate our method of including the LHC constraints, a scan with a likelihood that we derived
  from the \alphaTexp, and a likelihood from the experiments that
  constrain the Standard Model's nuisance parameters only.

\item To examine the impact of the current LHC constraints, a scan with a
  likelihood from non-LHC experiments and from the \alphaT\ but
  without a likelihood from \xenon.

\item To see the additional impact of the current \xenon\ limit scan
  with a likelihood from non-LHC experiments, the \alphaT\ and from
  \xenon.
\end{enumerate}

%
\subsection{\label{subsec:alphaT_cuts}The efficiency maps for the CMS \alphaT\ $1.1\invfb$ analysis}
We derived our LHC likelihood for the \cms\ search\cite{CMS:2011} for
$R$-parity conserving supersymmetry in all-hadronic events via a
kinematic variable \alphaT.  The results based on the LHC data sample
of $1.1\invfb$ of integrated luminosity recorded at $\sqrt{s} = 7\TeV$
showed no excess of events over the SM predictions. Our
aim was to translate the analysis scheme into a simplified approach to
obtain the signal selection efficiency for a large number of points in
the CMSSM parameter space.  To this end we generated a map of points
for the CMSSM parameters \mzero\ in the range of $(50,2000)\GeV$ and
\mhalf\ in the range of $(50,1000)\GeV$, both with a step of
$50\GeV$. We fixed the values of the other CMSSM parameters: $\azero =
0$, $\tanb = 10$ and $\signmu = 1$.  For each point, for a defined set
of the CMSSM parameters we calculated a mass spectrum and a table of
supersymmetric particle decays using the programs
\softsusy\cite{Allanach:2001kg} and \susyhit\cite{Djouadi:2006bz},
respectively. For each point, we then generated 10,000 events with the
Monte Carlo generator \pythia\cite{Sjostrand:2000wi} with a cross
section value obtained at the leading order. We analyzed events at the
generator level. Only the geometrical acceptance was applied to
simulate the \cms\ detector response.  Leptons were accepted in the
pseudorapidity range $\eta < 2.5$ for electrons and $\eta \le 2.1$
for muons, respectively. The isolation of leptons was checked at the
generator level. Stable, generator level particles, excluding
neutrinos, were clustered into jets using the anti-$k_T$
algorithm\cite{Cacciari:2008gp}, with a cone size parameter $R=0.5$.
Jets were accepted up to $|\eta|<3$.

The aim of the \alphaT\ analysis was to select hadronic events with
high transverse momentum jets. Initially, events with at least one jet
with transverse momentum $\pt > 50\GeV$ and $\eta > 3$ were accepted
if no isolated lepton or photon were found in the event, namely events
with an isolated lepton (electron or muon) with $\pt > 10\GeV$ or an
isolated photon with $\pt > 25\GeV$ were rejected.
Events had to satisfy a condition based on an \Ht\ variable, defined as 
${H_T = \sum_{i=1}^{n}E_T^{jet_i}}$; \Ht\ was required to be above $275\GeV$. 
Following the \cms\ analysis\cite{CMS:2011}, the trigger was fully efficient for selected 
events, and therefore no attempt to emulate the trigger was made.

The offline analysis used the following
\Ht\ binning: 
275 -- 325, 325 -- 375, 375 -- 475, 475 -- 575, 575 -- 675,
675 -- 775, 775 -- 875, and $>875\GeV$.
In each \Ht\ bin, we applied the same cuts on transverse momentum of
jets (leading, second, others) in the event as in
\refref{CMS:2011}. Details are shown in \reftable{tab:htbins}.

The main discriminator against QCD multijet production
is the variable \alphaT\ defined for dijet events as
\begin{equation}
\alpha_T = \frac{ E_T^{ jet_2}} { M_T}
= \frac{E_T^{ jet_2}}
        { \sqrt {(\sum_{i=1}^{2} E_T^{jet_i}  )^2 - (\sum_{i=1}^{2}
p_x^{jet_i}  )^2  - (\sum_{i=1}^{2} p_y^{jet_i}  )^2 } },
\end{equation}
where $E_T^{jet_2}$ is the transverse energy of the less energetic jet in the event with 
two jets and $M_T$ is a transverse mass of the dijet system defined above.

In events with more than two jets, two pseudojets were formed following the same 
strategy as in \refref{CMS:2011} in such a way that the \et\ difference between two pseudojets 
was minimised. The value of \et\ of each of the two pseudojets was obtained by a scalar 
summing of the contributing $E_T^{jet}$ of jets. 
In the ideal case, the dijet system had $E_T^{jet_1} = E_T^{jet_2}$ and jets are back-to-back 
which resulted in a limit of $\alphaT = 0.5$, where the momentum of jets is large compared 
to the masses of jets. For back-to-back jets with $E_T^{jet_1} \neq E_T^{jet_2}$, values of 
\alphaT\ were smaller than $0.5$. Signal events with missing transverse energy resulted in 
\alphaT\ greater than $0.5$. Therefore the QCD background was efficiently rejected with the 
final cut of $\alphaT > 0.55$. The remaining events after all cuts were compared with SM 
background expectation predictions. No excess was found and the numbers listed in 
\reftable{tab:htbins} were used as numbers of observed background events.

\begin{figure}
\centering
\subfloat[][The case of $275\GeV < H_T < 325\GeV$.]{%
\label{fig:Map-a}%
\includegraphics[width=0.49\linewidth]{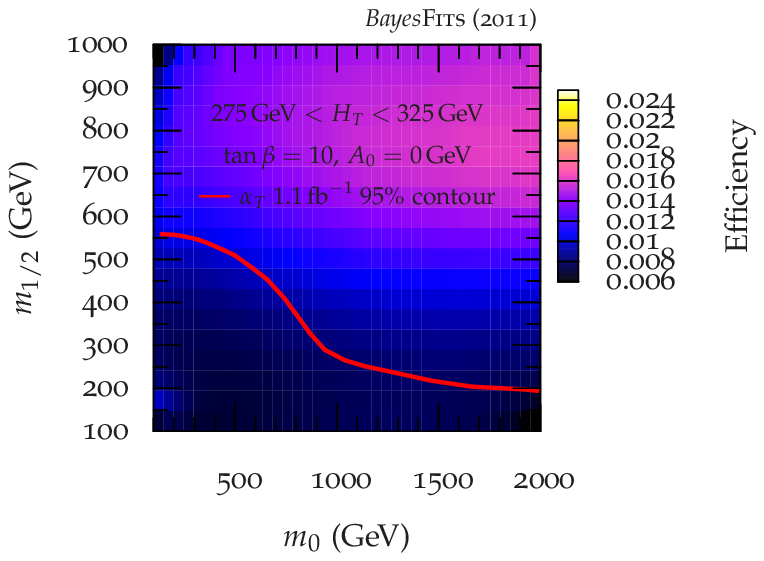}}%
\hspace{8pt}%
\subfloat[][The case of $325\GeV < H_T < 375\GeV$.]{%
\label{fig:Map-b}%
\includegraphics[width=0.49\linewidth]{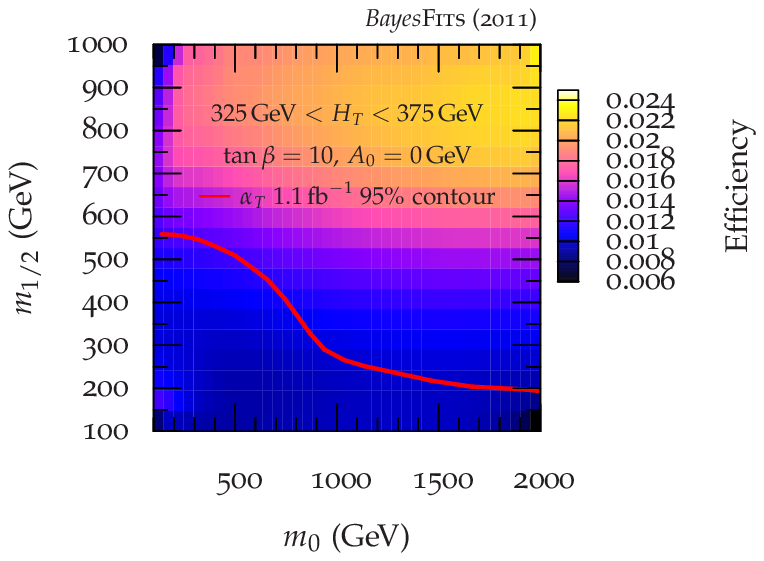}}
\caption[]{The efficiency maps of our approximation to the 
  \alphaTexp\ on the (\mzero, \mhalf) plane, for $\tanb = 10$ and
  $\azero = 0$, for the two dominant $H_T$ bins.} 
\label{fig:Map}
\end{figure} 

For the selection defined above, we prepared eight efficiency maps,
one for each \Ht\ bin, two of which are shown as examples in
\reffigure{fig:Map}. 

\begin{table}
\begin{tabular}{|l| c|c|c|c| c|c|c|c|}
\hline 
$H_T$ Bin (GeV)  & $275-325$ & $325-375$ & $375-475$ & $475-575$  & $575-675$ & $675-775$ & $775-875$ & $>875$\\
\hline 
$p_T^{\rm leading}$ (GeV)& $73$ & $87$ & $100$ & $100$ & $100$ & $100$ & $100$ & $100$ \\
$p_T^{\rm second}$ (GeV)& $73$ & $87$ & $100$ & $100$ & $100$ & $100$ & $100$ & $100$ \\ 
$p_T^{\rm others}$ (GeV)& $37$ & $43$ & $50$  & $50$  & $50$  & $50$  & $50$  & $50$ \\ 
\hline 
\multicolumn{9}{|c|}{Observed events} \\
\hline 
$\alpha_T>0.55$ & $782$ & $321$ & $196$ & $62$ & $21$ & $6$  & $3$ & $1$ \\
\hline 
\end{tabular}
\caption{Definition of the \Ht\ bins and the corresponding \pt\ thresholds for the leading, 
second, and all other remaining jets in the event. Observed events refers to the number 
of events passing  all \alphaT\ cuts for $1.1\invfb$ of data collected by the \cms\ 
Collaboration\cite{CMS:2011}.}
\label{tab:htbins}
\end{table}

Experimental selections involving missing transverse energy estimates and selections from an
analysis of the distance between a jet and \Ht\ were not implemented, because the relevant variables 
calculated at the generator level are not reliable. Nevertheless, as discussed below, good 
agreement between efficiency maps obtained in this analysis, and the \cms\ result shows that 
such an approximation is justified.

\subsection{\label{subsec:alphaT_like}The likelihood from the CMS \alphaT $1.1\invfb$ analysis}
The approximate efficiency maps derived in
\refsection{subsec:alphaT_cuts} for the \alphaTlim\ allow us to
evaluate a likelihood, so that we can find the regions of the CMSSM's
parameter space in best agreement with the currently strongest LHC
limit.

We assume that the experiment can be described by a Poisson
distribution; that is, that the events were independent and that the
likelihood of observing events was described by a Poisson
distribution.  We assume that the total number of observed events was
the sum of contributions from supersymmetric processes and from SM
processes. Following the $\alpha_T$ cuts described in
\refsection{subsec:alphaT_cuts}, we consider separately the eight bins
of the kinematical variable $H_T$ summarized in \reftable{tab:htbins}.

The number of supersymmetric events that we expected in an $i$-th
$H_T$ bin, $s_i$, is the product of the detector efficiency for that
bin (the fraction of events that survive the 
\alphaT\ cuts), $\epsilon_i$, the integrated luminosity,
$\int L=1.1\invfb$, and the total cross section for the production of supersymmetric
particles at $\sqrt{s}=7\TeV$, $\sigma$,
\begin{equation}\label{eqn:signal}
 s_i = \epsilon_i \times \sigma \times \int L.  
\end{equation}

The likelihood for this case, $\mathcal{L}$, which is the probability of observing a
set of $\{o_i\}$ events given that we expected $\{s_i\}$ 
supersymmetric events and $\{b_i\}$ SM background events, 
is a product of Poisson distribution for each bin with means $\lambda_i = s_i + b_i$,
\begin{equation}
\label{eqn:likelihood}
 \mathcal{L} = \prod_{i} \frac{e^{-(s_i + b_i)}\left(s_i + b_i \right)^{o_i}}{o_i !}. 
\end{equation}
We assume that Standard Model backgrounds, $b_i$, are precisely
known. If their experimental errors were significant, the expression
for the likelihood would be multiplied by distributions describing the
SM backgrounds\cite{Loparco:2011yr} and they would be included as
nuisance parameters. This is not the case for the \alphaT\ analysis.

\begin{figure}
\centering
\subfloat[][The likelihood map from our \alphaTexp\ on the (\mzero,
\mhalf) plane, for $\tanb = 10$ and $\azero = 0$ 
using grid scan.]{%
\label{fig:LikeMap-a}%
\includegraphics[scale=1.3065]{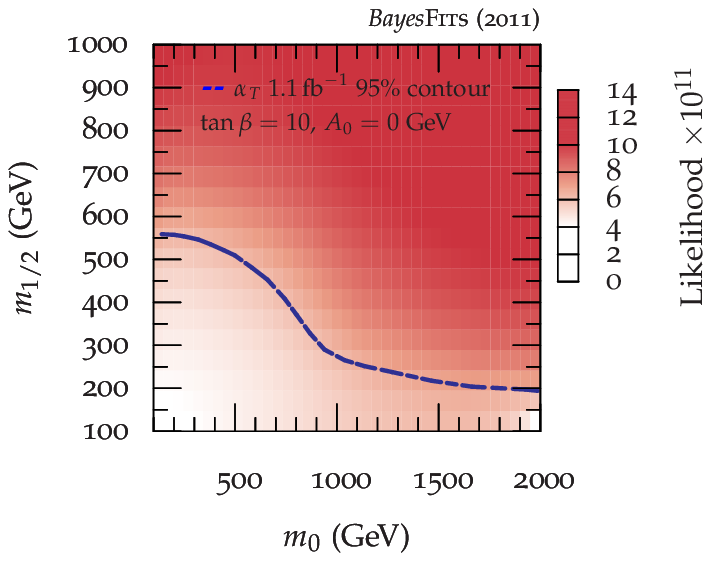}}
\hspace{8pt}%
\subfloat[][Profile likelihood regions on the (\mzero, \mhalf) plane, with
a likelihood from our \alphaTexp, using Nested Sampling and a log prior.]{%
\label{fig:LikeMap-b}%
\raisebox{0.37cm}{\includegraphics{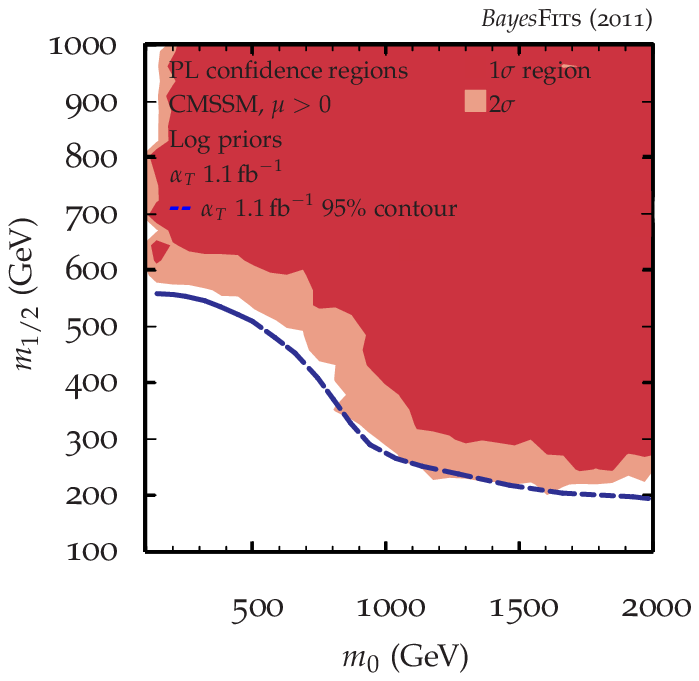}}}%
\caption[]{The maps of (a) the likelihood and (b) the profile
  likelihood for our approximation to the \alphaTexp\ on the (\mzero, \mhalf)
  plane. The dashed blue lines show the official \cms\ $95\%$ exclusion
  contour.}
\label{fig:LikeMap}
\end{figure} 

All-hadronic final-state processes are expected to be independent of
\tanb\ and \azero, because these parameters have little influence on
the squark and gluino masses, and this has been confirmed by the CMS
Collaboration in the \alphaTexp, as well as, \eg, in
\refref{Allanach:2011ut,Bechtle:2011dm}, and also in this study.  For
this reason we created efficiency maps using the mentioned earlier
fixed-grid scan with fixed values of $\tanb = 10$ and $\azero=0$, as
described in \refsection{subsec:alphaT_cuts}.
The cross section was calculated to leading order with
\pythia\cite{Sjostrand:2000wi}.\footnote{The cross section, and
  consequently the number of expected supersymmetric events, changes
  by over 10 orders of magnitude over the (\mzero,\mhalf) plane. The
  resulting likelihood function is not, therefore, sensitive to
  next-to-leading order corrections to the cross section. Even if
  $\sigma_\text{NLO} \sim \sigma_\text{LO}$, the corrections would
  only slightly shift the isocontours of the cross section and likelihood
  on the (\mzero, \mhalf) plane.}
We used this information to produce a
likelihood map using \refequation{eqn:signal} and
\refequation{eqn:likelihood}. 

The resulting likelihood map is shown in \reffigure{fig:LikeMap-a}.
It exhibits the correct behavior; below the official
$95$\% contour obtained by \alphaTexp\cite{CMS:2011}, which is also
indicated, the likelihood is small. Our approximate $95$\% contour was
calculated with the help of \refequation{eqn:conf_interval}. It
corresponds to the boundary of the $2\sigma$ range of the profile
likelihood 
obtained using Nested Sampling and a log prior, which
is 
shown in \reffigure{fig:LikeMap-b}, and is very close to
the official \cms\ contour which is also shown in the Figure. For each
CMSSM point for which our scanning algorithm wished to evaluate the
likelihood, we interpolated the likelihood from our likelihood map.

\section{\label{sec:results}Results}
In this section we will present our numerical results. To start with,
in Table~\ref{tab:priors} we show prior ranges and distributions of
CMSSM parameters and of SM nuisance parameters, while in
Table~\ref{tab:exp_constraints} we list the observables that we will
use in our analysis. These come in three sets. First, by ``Non-LHC''
we collectively denote all relevant constraints from dark matter,
\abundchi, precision measurements: \sineff, \mw, \etc, flavour
physics: \brbsg, \brbtn, \deltam, and \brbsmumu,\footnote{Actually, LHCb\cite{Aaij:2011rja}, an LHC
  experiment, recently obtained the  best upper limit of
  $\brbsmumu<1.5 \times 10^{-8}$, but in this  study we will nevertheless include it in the
  ``Non-LHC'' group of constraints.} the excess in the anomalous magnetic moment of
the muon, \deltaa, as
well as LEP and Tevatron limits on the Higgs sector and superpartner
masses. Second, we will apply currently the best limit
published by the \cms\ Collaboration, which is from its \alphaT\ analysis of 1.1\invfb\
of data. Finally, at the end we will also apply an upper limit on the elastic
scattering of neutralino dark matter \sigsip\ recently published by
\xenon, in order to investigate its additional impact on
the CMSSM's parameters and various observables.

\begin{table}
\begin{tabular}{|l|l|l|l|l|l|}
\hline 
Measurement & Mean & Exp.~error & The.~error & Likelihood distribution & Reference\\
\hline 
\multicolumn{6}{|l|}{\alphaTexp} \\
\hline 
\alphaT  	& See text 	& See text  	& $0$ 	& Poisson  &\cite{CMS:2011}\\
\hline 
\multicolumn{6}{|l|}{\xenon} \\
\hline 
$\sigsip\left(m_\neutralinoone \right)$	 & $< f\left(m_\neutralinoone \right)$, see text 	& $0$ 	& $1000\%$ 	& Upper limit: error function  &\cite{Aprile:2011dd}\\
\hline 
\multicolumn{6}{|l|}{Non-LHC} \\
\hline 
\abundchi 			& $0.1120$ 	& $0.0056$  	& $10\%$ 		& Gaussian &  \cite{Komatsu:2010fb}\\
\sineff 			& $0.231\,16$   & $0.000\,13$   & $0.000\,15$           & Gaussian &  \cite{Nakamura:2010zzi}\\
\mw                     	& $80.399$      & $0.023$   	& $0.015$               & Gaussian &  \cite{Nakamura:2010zzi}\\
\deltaa $\times 10^{10}$ 	& $30.5 $  	& $8.6$ 	& $1.0$ 		& Gaussian &  \cite{Nakamura:2010zzi,Miller:2007kk} \\
\brbsg $\times 10^{4}$ 		& $3.60$   	& $0.23$ 	& $0.21$ 		& Gaussian &  \cite{Nakamura:2010zzi}\\
\brbtn $\times 10^{4}$          & $1.66$  	& $0.66$ 	& $0.38$ 		& Gaussian &  \cite{Asner:2010qj}\\
\deltam                         & $17.77$ 	& $0.12$ 	& $2.40$  		& Gaussian &  \cite{Nakamura:2010zzi}\\
\brbsmumu			& $< 1.5 \times 10^{-8}$  	& $0$ & $14\%$    	& Upper limit: error function &  \cite{Aaij:2011rja}\\
\hline 
\multicolumn{6}{|l|}{Nuisance} \\
\hline 
\alphaemmz	 		& $127.916$ & $0.015$ & $0$ & Gaussian &  \cite{Nakamura:2010zzi}\\
\mt             		& $172.9$   & $1.1$   & $0$ & Gaussian &  \cite{Nakamura:2010zzi}\\
\mb                  		& $4.19$    & $0.12$  & $0$ & Gaussian &  \cite{Nakamura:2010zzi}\\
\alphas               		& $0.1184$  & $0.0006$& $0$ & Gaussian &  \cite{Nakamura:2010zzi}\\
\hline 
\multicolumn{6}{|l|}{LEP and Tevatron $95$\% limits} \\
\hline 
$m_h$          		& $>114.4$   		 & $0$& $3$   & Lower limit: error function & \cite{Barate:2003sz} \\
$\zeta_h^2$      	& $<f\left(m_h\right)$	 & $0$& $0$   & Upper limit -- step function  & \cite{Barate:2003sz} \\
$m_{\chi}$   		& $>50$      		 & $0$& $5$\% & Lower limit: error function & \cite{Heister:2003zk} (\cite{Heister:2001nk,*Achard:2003ge}) \\
$m_{\chi^\pm_1}$	& $>103.5$  ($92.4$)     & $0$& $5$\% & Lower limit: error function & \cite{lepsusy} (\cite{Heister:2001nk,*Achard:2003ge}) \\
$m_{\tilde{e}_R}$ 	& $>100$    ($73$)       & $0$& $5$\% & Lower limit: error function & \cite{lepsusy} (\cite{Heister:2001nk,*Achard:2003ge})  \\
$m_{\tilde{\mu}_R}$ 	& $>95$     ($73$)  	 & $0$& $5$\% & Lower limit: error function & \cite{lepsusy} (\cite{Heister:2001nk,*Achard:2003ge})  \\
$m_{\tilde{\tau}_1}$ 	& $>87$     ($73$)  	 & $0$& $5$\% & Lower limit: error function & \cite{lepsusy} (\cite{Heister:2001nk,*Achard:2003ge}) \\
$m_{\tilde{\nu}}$ 	& $>94$     ($43$)  	 & $0$& $5$\% & Lower limit: error function & \cite{Abdallah:2003xe} (\cite{Nakamura:2010zzi}) \\
$m_{\tilde{t}_1}$ 	& $>95$	    ($65$)  	 & $0$& $5$\% & Lower limit: error function & \cite{lepsusy} (\cite{Heister:2001nk})\\
$m_{\tilde{b}_1}$ 	& $>95$	    ($59$)  	 & $0$& $5$\% & Lower limit: error function & \cite{lepsusy} (\cite{Heister:2001nk})\\
$m_{\tilde{q}}$ 	& $>375$    		 & $0$& $5$\% & Lower limit: error function & \cite{Abazov:2009zi} \\
${\mgluino}$  	& $>289$     		 & $0$& $5$\% & Lower limit: error function & \cite{Abazov:2009zi} \\ 
\hline 
\end{tabular}\caption{The experimental measurements that constrain the CMSSM's
  parameters and the Standard Model's nuisance parameters. Masses are
  in GeV. The numbers in parentheses in the list of LEP and Tevatron experimental
  measurements are weaker experimental bounds, which we use for some
  sparticle mass hierarchies.} 
\label{tab:exp_constraints}
\end{table}

Below we will present the results of our scans as one-dimensional (1D)
or two-dimensional (2D) marginalized posterior pdf maps
of the CMSSM's parameters and observables. In evaluating the
posterior pdfs, we marginalized over the CMSSM's other parameters and
the SM's nuisance parameters, 

\subsection{\label{sec:alphatimpact}Impact of the \alphaT\ limit}
\begin{figure}
\centering
\subfloat[][The CMSSM's parameters constrained by the non-LHC
experiments only.]{%
\label{fig:Comparison_m0m12_lhc-a}%
\includegraphics[scale=\scale]{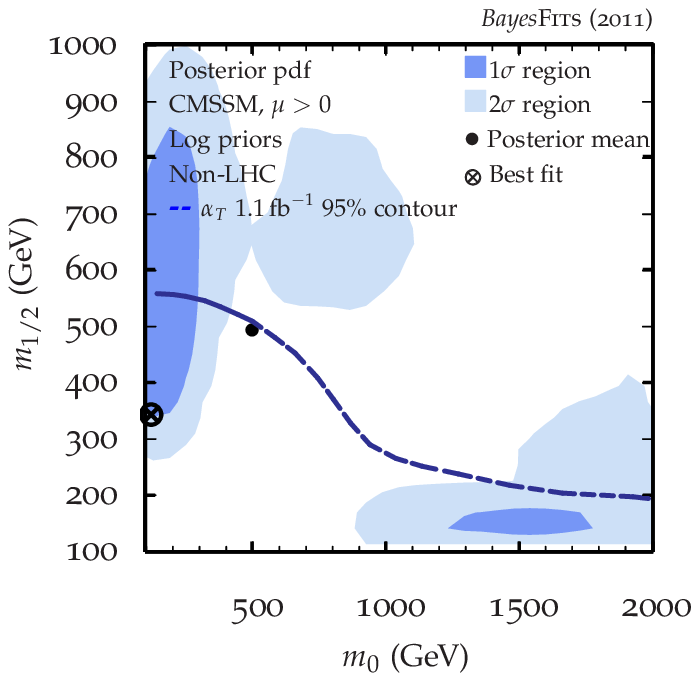}}%
\hspace{8pt}%
\subfloat[][The CMSSM's parameters constrained by the \alphaTexp\ and by the non-LHC experiments.]{%
\label{fig:Comparison_m0m12_lhc-b}%
\includegraphics[scale=\scale]{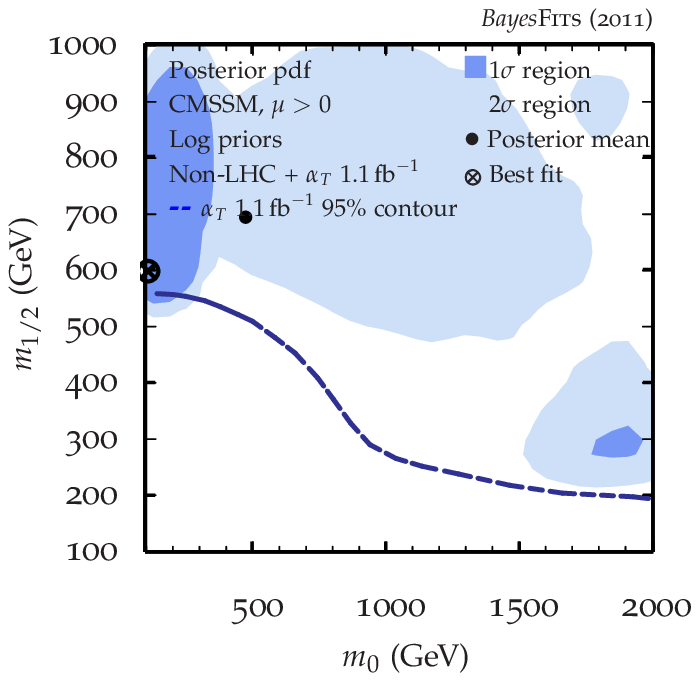}}
\caption[]{Marginalised posterior pdf on the (\mzero, \mhalf) plane,
  \subref{fig:Comparison_m0m12_lhc-a} before and
  \subref{fig:Comparison_m0m12_lhc-b} after we included a likelihood from the \alphaTlim.}
\label{fig:Comparison_m0m12_lhc}
\end{figure}

First, in \reffigure{fig:Comparison_m0m12_lhc}, on the (\mzero,
\mhalf) plane, we show the results of a scan with a likelihood from
the non-LHC constraints (left panel) and with the additional impact of
imposing the \alphaT\ $1.1\invfb$ constraint (right panel).  Dark
(light) blue regions denote the $1\sigma$ ($2\sigma$) posterior pdf
regions of the (\mzero, \mhalf) plane.  In
\reffigure{fig:Comparison_m0m12_lhc-a} one can see two distinct modes
on the (\mzero, \mhalf) plane. The vertical $1\sigma$ mode is located in in the
stau coannihilation/$\ha$--funnel (SC/AF) region, while  the
associated $2\sigma$ region is caused by the AF only.
On the opposite
side of the diagonal of the (\mzero, \mhalf) plane, the horizontal
mode corresponds to overlapping contributions from the light Higgs resonance region (the
narrow $1\sigma$ strip) and, above it, the $2\sigma$ focus point (FP)/horizontal branch (HB) region.  

As a result, the posterior mean (denoted by a solid black
dot) lies between the two modes, but closer to the SC/AF region, where
the best-fit point (denoted by an encircled cross) is also located.
The blue dashed line denoting the $95\%$ lower limit derived by the
\alphaTexp\ is marked, but not applied here.

The existence of the two broad regions is consistent with the findings of
previous pre-LHC Bayesian
analyzes\cite{deAustri:2006pe,Roszkowski:2007fd,Feroz:2008wr,Trotta:2008bp},
although their relative size does depend on the choice of the
prior. Physically they both arise as a result of reducing the relic
density of the neutralino, which is generally too large in the
part of the (\mzero, \mhalf) plane where \neutone\ is the lightest
superpartner. 
In the SC
region the density is reduced by a coannihilation with the
lighter stau (and other sleptons), and likewise in the AF region it is
reduced by
neutralino pair annihilation through the pseudoscalar Higgs
$\ha$--funnel, hence the name. Likewise, the same mechanism is at play
in the narrow horizontal  light Higgs resonance region.
In the FP
region, on the other hand, as one moves down along a line roughly perpendicular to the
diagonal of the (\mzero, \mhalf) plane, the value of $\mu^2$ drops
down from large values (for which the Lightest Supersymmetric Particle (LSP) is very binolike, with too
large \abundchi), and eventually becomes negative, implying a failure
of radiative electroweak symmetry breaking conditions. Close to this
boundary, in a rather narrow strip of the plane, $\mu$ is of order
\mhalf, the LSP develops a sizable Higgsino component, and the relic
abundance becomes acceptable.

The impact of the \alphaTlim, which is implemented in our analysis by
simulating detector efficiency and evaluating the corresponding
likelihood, as described in \refsection{subsec:alphaT_like}, is shown
in \reffigure{fig:Comparison_m0m12_lhc-b}.  Clearly, the \alphaT\
limit has cut deep into the (\mzero, \mhalf) plane's high posterior
probability regions -- a nominal fraction of the posterior pdf $
1\sigma$ region is outside of the \alphaT\ 95\% confidence
interval. The \alphaT\ limit pushed the credible regions, as well as
the best-fit point, to larger values of \mhalf.  Significantly, the
two modes on the (\mzero, \mhalf) plane have remained. The
$1\sigma$ SC/AF region has been pushed up nearly vertically, while
the $2\sigma$ one of AF only has become inflated and extended to larger values of
(\mzero, \mhalf). 
Note also that the horizontal light Higgs resonance region
has now completely disappeared.
On the other hand, the FP/HB region is not excluded
by the \alphaTlim; rather, it is pushed to larger values of \mzero\
and, to a lesser extent, \mhalf.

\begin{figure}
\centering
\subfloat[][The CMSSM's parameters constrained by the non-LHC experiments.]{%
\label{fig:Comparison_a0tanb_lhc-a}%
\includegraphics[scale=\scale]{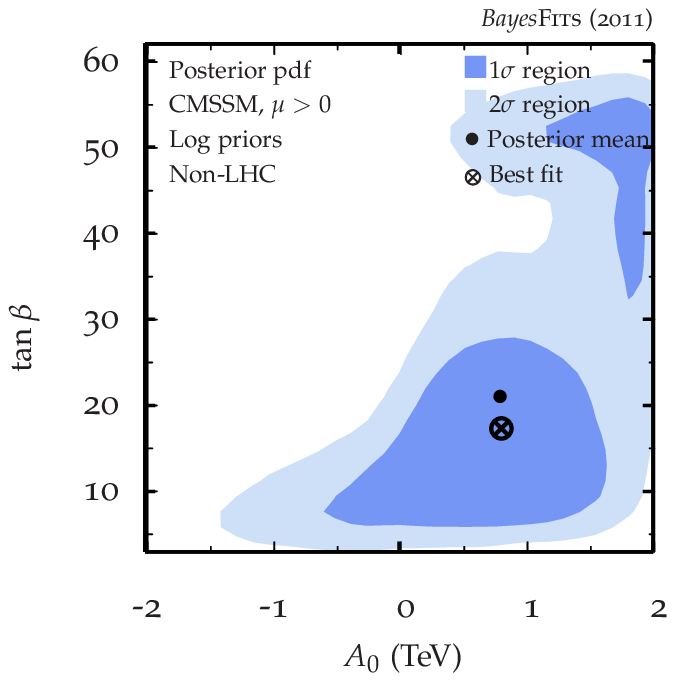}}%
\hspace{8pt}%
\subfloat[][The CMSSM's parameters constrained by the \alphaTexp\ and non-LHC experiments.]{%
\label{fig:Comparison_a0tanb_lhc-b}%
\includegraphics[scale=\scale]{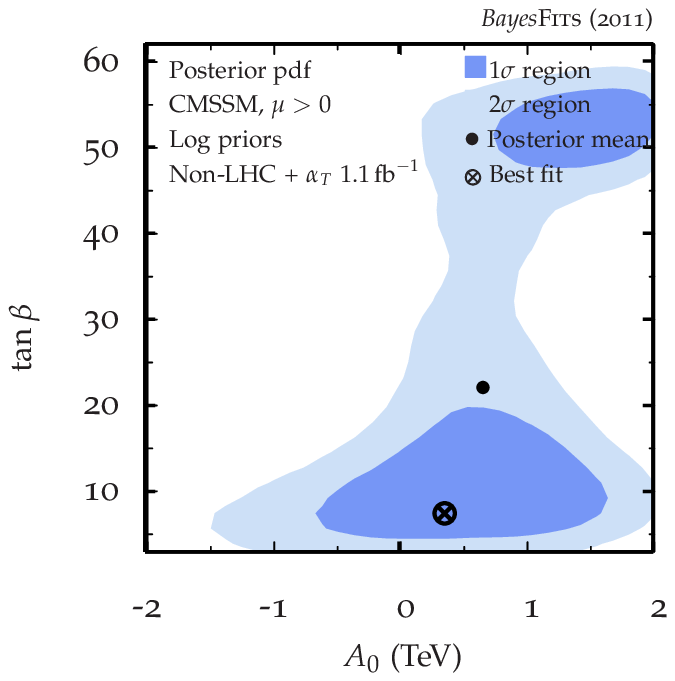}}
\caption[]{Marginalised posterior pdf on the (\azero, \tanb) plane, \subref{fig:Comparison_a0tanb_lhc-a} before and  \subref{fig:Comparison_a0tanb_lhc-b} after we included a likelihood from the \alphaTlim.}
\label{fig:Comparison_a0tanb_lhc}
\end{figure}

An analogous comparison of the pre- and post-\alphaT\ situation in the
(\azero, \tanb) plane is presented in
\reffigure{fig:Comparison_a0tanb_lhc-a} and
\reffigure{fig:Comparison_a0tanb_lhc-b}.  There are again two
$1\sigma$ modes: one at relatively small values of $\tanb \lsim 20$
and the other one at much larger values ($\sim55$), although at
$2\sigma$ an entire range of scanned values of \tanb\ is allowed. The
first mode corresponds to the $1\sigma$ stau coannihilation region in
\reffigure{fig:Comparison_m0m12_lhc-b}; however, in the much larger,
$2\sigma$, mostly $\ha$-funnel region of the (\mzero,\mhalf) plane
much larger values of $\tanb \sim 55$ are predominant. On the other
hand, in in the focus point region of large \mzero\ in
\reffigure{fig:Comparison_m0m12_lhc-b} we find $25\lsim\tanb\lsim
55$.

We can see that the application of the \alphaT\ constraint narrows
down both modes in the posterior pdf on the (\azero, \tanb) plane,
primarily as a result of pushing up and out the focus point region,
with the effect of strongly disfavoring midrange values of \tanb. On
the other hand the \cms\ limit shows fairly little effect on \azero,
which remains poorly determined.

Although our \alphaT\ likelihood is independent of \tanb\ and \azero,
it is clear from \reffigure{fig:Comparison_a0tanb_lhc} that \alphaT\
does impact these parameters. The CMSSM's parameters are correlated in
a nontrivial way; once the \alphaT\ limit is added to the likelihood it alters
the best-fitting values of \mzero\ and \mhalf\, and the values of
\tanb\ and \azero\ must be adjusted accordingly, so that the predicted
values for the CMSSM's observables maintain agreement with the
experimental measurements.

\begin{figure}
\centering
\subfloat[][]{%
\hspace{8pt}%
\label{fig:NonLHC+alphaT_scan-a}%
\includegraphics[scale=\scale]{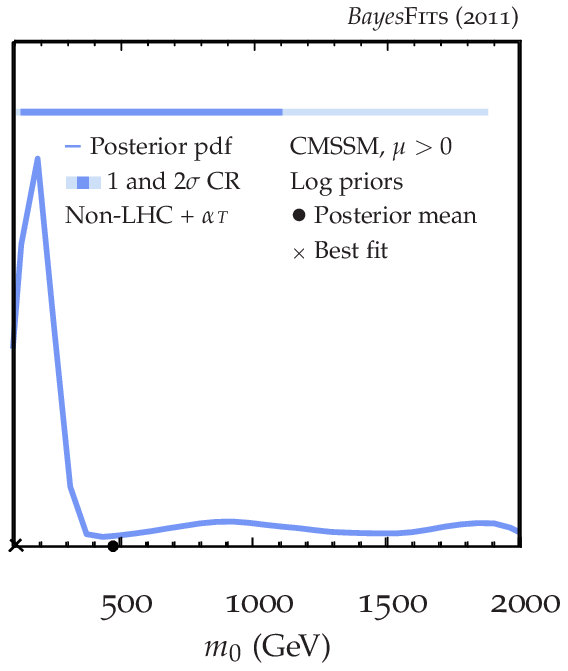}}
\hspace{8pt}%
\subfloat[][]{%
\label{fig:NonLHC+alphaT_scan-b}%
\includegraphics[scale=\scale]{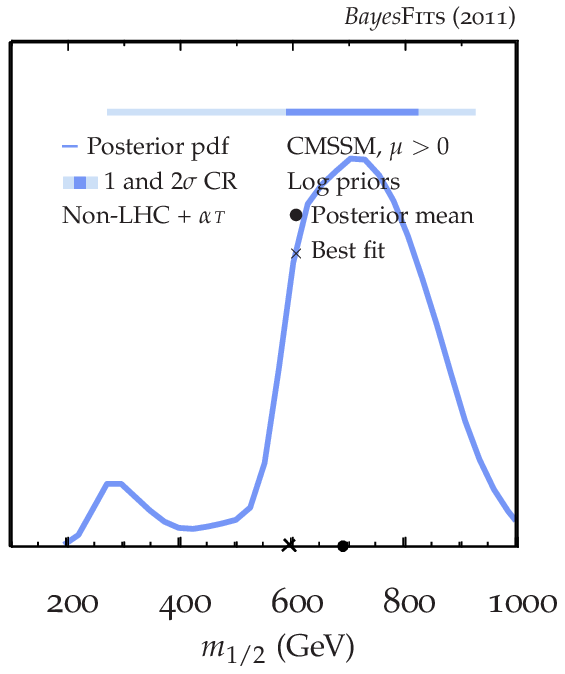}}\\
\subfloat[][]{%
\label{fig:NonLHC+alphaT_scan-c}%
\includegraphics[scale=\scale]{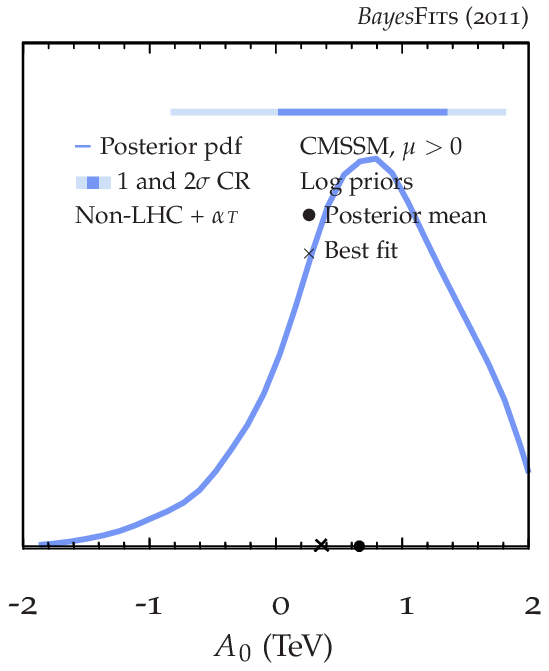}}%
\hspace{8pt}%
\subfloat[][]{%
\label{fig:NonLHC+alphaT_scan-d}%
\includegraphics[scale=\scale]{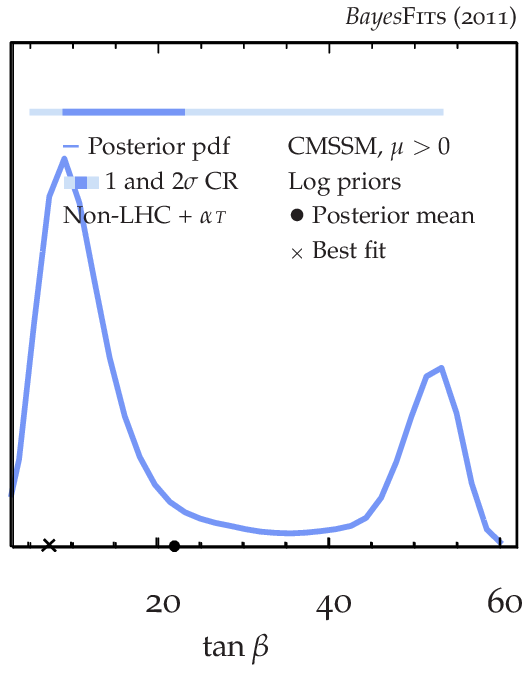}}%
\caption[]{One-dimensional marginalized posterior pdf 
for the CMSSM's parameters constrained by the \alphaTexp\
  and by the non-LHC experiments. 
   The dark (light) blue horizontal bars span the one-
  (two-)dimensional central credible regions.
} 

\label{fig:NonLHC+alphaT_scan}
\end{figure}

In \reffigure{fig:NonLHC+alphaT_scan} we show 1D marginalized
posterior pdf 
plots for the CMSSM's parameters,
constrained by the non-LHC experiments and by the \alphaT\ limit. We
also show the
dark (light) blue horizontal bars which indicate the one- (two-) dimensional
central credible regions.
The high
probability modes in the \mzero\ distributions at $\mzero \sim
200\GeV$ in \reffigure{fig:NonLHC+alphaT_scan-a} and in the \mhalf\
distribution at $\mhalf \sim 700\GeV$ in
\reffigure{fig:NonLHC+alphaT_scan-b} correspond to the high
probability mode in the SC/AF region in
\reffigure{fig:Comparison_m0m12_lhc-b}. \reffigure{fig:NonLHC+alphaT_scan-c}
shows that the experimental constraints favor positive values of
\azero. The two modes in the \tanb\ distribution in
\reffigure{fig:NonLHC+alphaT_scan-d} correspond to the two modes in
\reffigure{fig:Comparison_a0tanb_lhc}.

After marginalisation, it is difficult to identify in the \mzero\ 1D
marginalized posterior pdf the FP/HB mode, which was
present in the 2D marginalized posterior pdf in
\reffigure{fig:Comparison_m0m12_lhc}, though it is present as a second
smaller mode in the \mhalf\ 1D marginalized posterior pdf in
\reffigure{fig:NonLHC+alphaT_scan-b}. This is caused by the
contribution of the large $2\sigma$ credible region at both large \mhalf\
and \mzero.

\begin{figure}
\centering
\subfloat[][]{%
\label{fig:4masses-a}%
\includegraphics{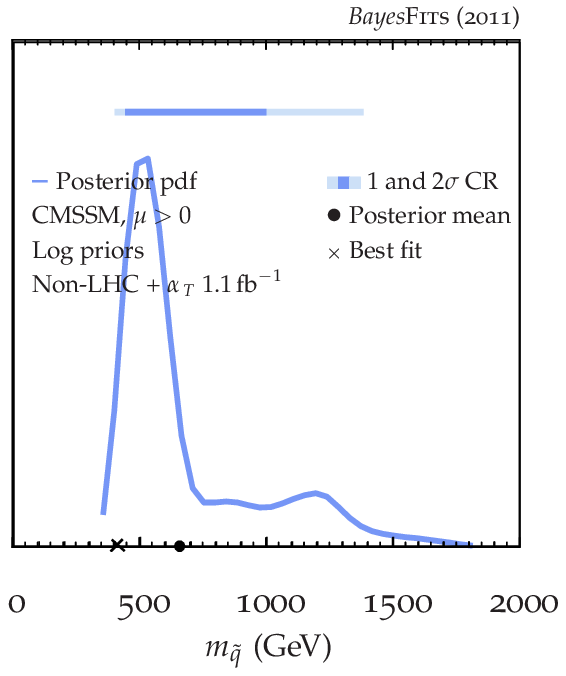}}%
\hspace{8pt}%
\subfloat[][]{%
\label{fig:4masses-b}%
\includegraphics{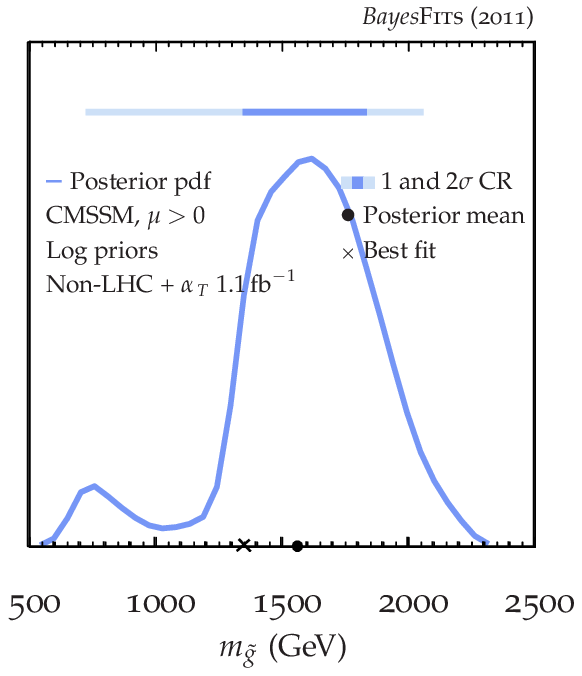}}\\
\subfloat[][]{%
\label{fig:4masses-c}%
\includegraphics{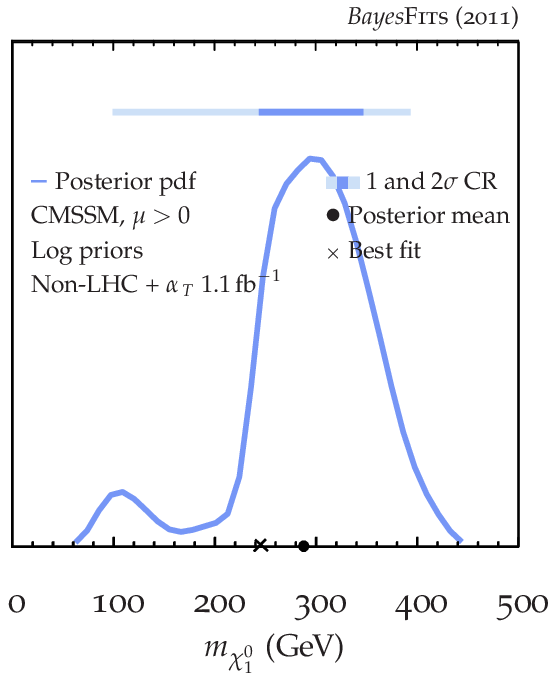}}%
\hspace{8pt}%
\subfloat[][]{%
\label{fig:4masses-d}%
\includegraphics{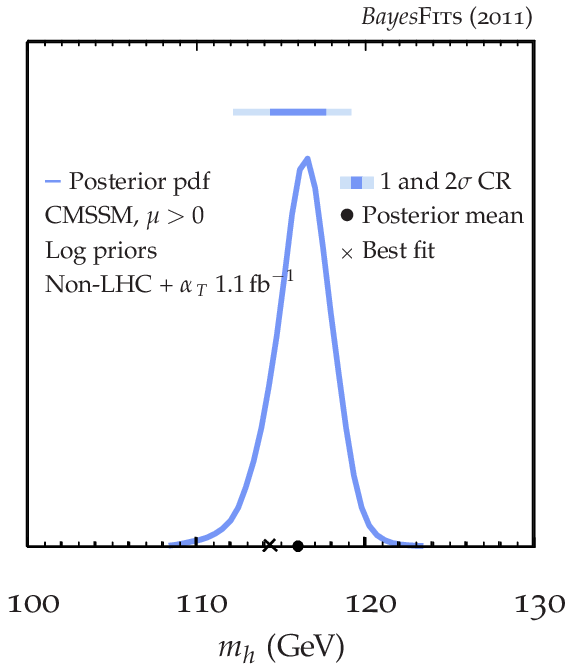}}%
\caption[]{The mass of the lightest \subref{fig:4masses-a} squark, \subref{fig:4masses-b} gluino, 
 \subref{fig:4masses-c} neutralino, and \subref{fig:4masses-d}
 Higgs-boson in the CMSSM constrained by \alphaT\ and by the non-LHC
 experiments. The horizontal bars have been defined in
 \protect\reffigure{fig:NonLHC+alphaT_scan}. }%
\label{fig:4masses}%
\end{figure}

In \reffigure{fig:4masses} we present 1D plots of
marginalized posterior pdf 
for notable particle
masses, when we scanned the CMSSM with a likelihood from \alphaT\ and
from the non-LHC experiments. These distributions are typically
bimodal; one mode is from the mode in the SC/AF region
and one mode is from the mode in the FP/HB region on the
(\mzero, \mhalf) plane in \reffigure{fig:Comparison_m0m12_lhc}.

\reffigure{fig:4masses-a} shows that the posterior 
favors a moderately heavy lightest squark\footnote{The mass
  of the lightest squark $m_{\squark} = \min
  \left(m_{\squark_i}\right)$} (the lighter stop) with $m_{\squark} \sim 500\GeV$,
corresponding to the mode in the SC/AF region with small $\mzero \sim
100\GeV$. There is a second, much smaller, mode in the posterior
at $m_{\squark} \sim
1250\GeV$, corresponding to the mode in the FP/HB region with large
$\mzero \sim 2000\GeV$. On the other hand, the squarks of the first
two generations can be much heavier, up to $\sim2.7\TeV$ at $1\sigma$.

Heavy gluinos, much heavier than lightest squarks, are favored with $\mgluino
\sim 1500\GeV$, as illustrated in \reffigure{fig:4masses-b}. This
dominant range corresponds to the mode in the SC/AF region with large
$\mhalf \sim 700\GeV$. Lighter gluinos with $\mgluino \alt 1000\GeV$ are
not, however, excluded, by the credible region.
This is a result of the mode in the FP/HB region.

\reffigure{fig:4masses-c} shows that 
the posterior pdf 
favors a lightest neutralino mass of
$m_\neutralinoone \sim 300\GeV$. This corresponds to a binolike
neutralino, with $m_\neutralinoone \sim 0.4 \mhalf$.

Lastly, \reffigure{fig:4masses-d} shows the 1D posterior pdf 
for the mass of the lightest Higgs in the
CMSSM, in agreement with pre-LHC results\cite{deAustri:2006pe,Allanach:2007qk,Trotta:2008bp,Buchmueller:2010ai}. 

The lightest Higgs boson in the CMSSM is to a very good approximation
SM like and the LEP limit ($114.4\GeV$) applies. Note, however, that
predicted Higgs masses below the LEP limit are 
permitted, because our likelihood function includes a $3\GeV$
theoretical error in the predicted Higgs mass. On the other side, the
1D posterior pdf is rather narrow, with the $95$\% credible region in
the range from $112.2\GeV$ to $119.2\GeV$ and the best-fit value of
$114.4\GeV$. 

\begin{figure}
\centering
\subfloat[][Non-LHC experiments.]{%
\label{fig:Comparison_mAtanbeta-a}%
\includegraphics[scale=\scale]{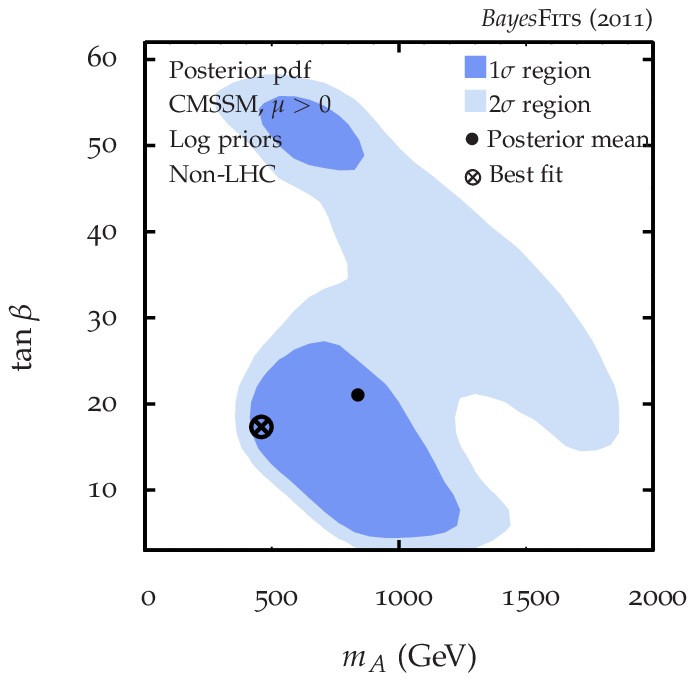}}%
\hspace{8pt}%
\subfloat[][\alphaTexp\ and the non-LHC experiments.]{%
\label{fig:Comparison_mAtanbeta-b}%
\includegraphics[scale=\scale]{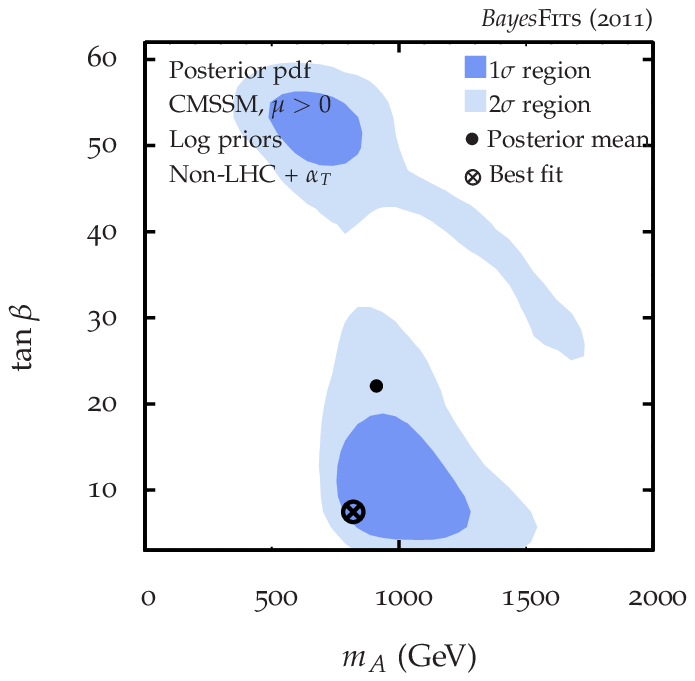}}
\caption[]{Marginalised posterior pdf on the ($m_A$, \tanb) plane, \subref{fig:Comparison_mAtanbeta-a} before and \subref{fig:Comparison_mAtanbeta-b} after we included a likelihood from the \alphaTlim.}
\label{fig:Comparison_mAtanbeta}
\end{figure}

Being SM like, the lightest Higgs boson may well be the only Higgs
state of the CMSSM accessible in Run~I.  The other Higgses are
typically nearly mass degenerate, and much heavier than the lightest
Higgs. This is illustrated in
\reffigure{fig:Comparison_mAtanbeta}, where we present a 2D posterior
pdf map on the ($m_A$, \tanb) plane in two cases: pre-\alphaT\ (left
panel) and post-\alphaT\ (right panel).
\reffigure{fig:Comparison_mAtanbeta} shows that the \alphaT\ limit favors a
heavier CP-odd neutral Higgs, because it favors higher values of
\mzero. Intermediate values of $\tanb \sim 30$, however, are
disfavored by \alphaT, and, consequently, very heavy CP-odd neutral
Higgs with $m_A \agt 1500\GeV$ become disfavored.

\subsection{\label{sec:xenonimpact}Impact of the \xenon\ limit}
The \xenon\ Collaboration has recently published a new $90$\%~C.L.
exclusion limit on the ($m_\neutone$, \sigsip) plane which significantly improves their previous
preliminary limit\cite{Aprile:2011hi}. In this section we
examine the impact of this new limit on the CMSSM's parameters and observables.

A proper implementation of the \xenon\ exclusion limit in our
likelihood function is somewhat tricky.  The $90$\% exclusion contour
alone is insufficient to reconstruct the likelihood function.  On top
of it, there are significant errors, which originate from poorly known
inputs which are needed in evaluating a limit on \sigsip\ from experimental data.
First, one assumes a ``default'' value for the local density of
$0.3\GeV/c^2$, but errors in astrophysical parameters which are
required to determine the local halo and its velocity distribution can
result in errors in \sigsip\ of order $2$\cite{Pato:2010zk}, although
a much better determination has been claimed\cite{Catena:2009mf}.
More significantly, the fractional error in evaluating \sigsip\ coming
from uncertainties in inputs to hadronic matrix elements can be of
order $5$, as discussed in detail in\cite{Buchmueller:2011ki}. The
main uncertainty is due to a poor knowledge of the $\pi$-nucleon
$\sigma$ term, $\Sigma_{\pi N}$, which determines the strange quark
component of the nucleon.

Given these large uncertainties, we neglect the experimental error in
the \xenon\ $90$\% exclusion contour and approximate it in the
likelihood with  a step function, 
$\mathcal{L}(\sigsip, m_\neutone) = 0 (1)$ for $\sigsip >(\leq)  \sigma^{\rm SI}_{p,\, 90\%} (m_\neutone)$,
where $\sigma^{\rm SI}_{p,\, 90\%} (m_\neutone)$ is the DM mass
dependent \xenon\ 90\% limit.  To incorporate the above errors in evaluating
\sigsip\ in a conservative way, we convolute the step function with a
Gaussian with $\mu = \sigsip$ and $\sigma = 10 \times \sigsip$,
resulting in a Gaussian error function. This will cause a rather
large smearing out of the \xenon\ limit.

\begin{figure}
\centering
\subfloat[][
]{%
\label{fig:Comparison_sigsipmchi-a}%
\includegraphics[scale=\scale]{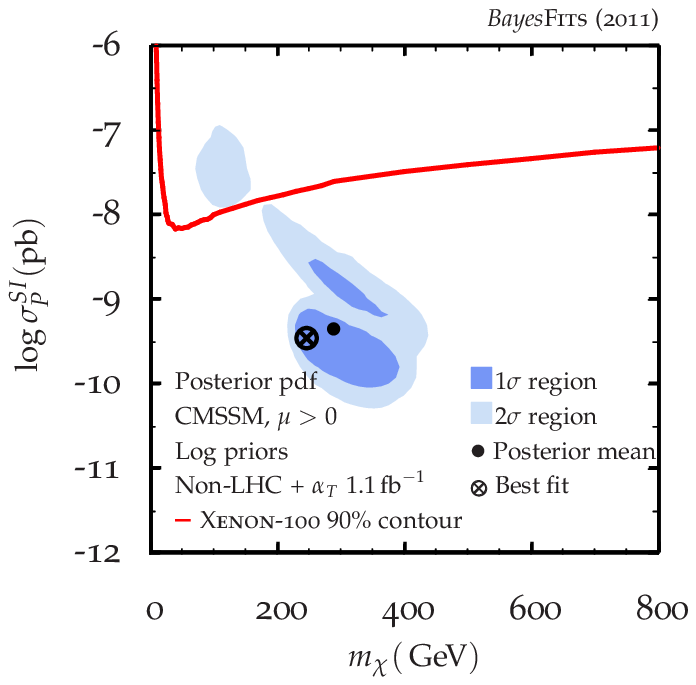}}%
\hspace{8pt}%
\subfloat[][
]{%
\label{fig:Comparison_sigsipmchi-b}%
\includegraphics[scale=\scale]{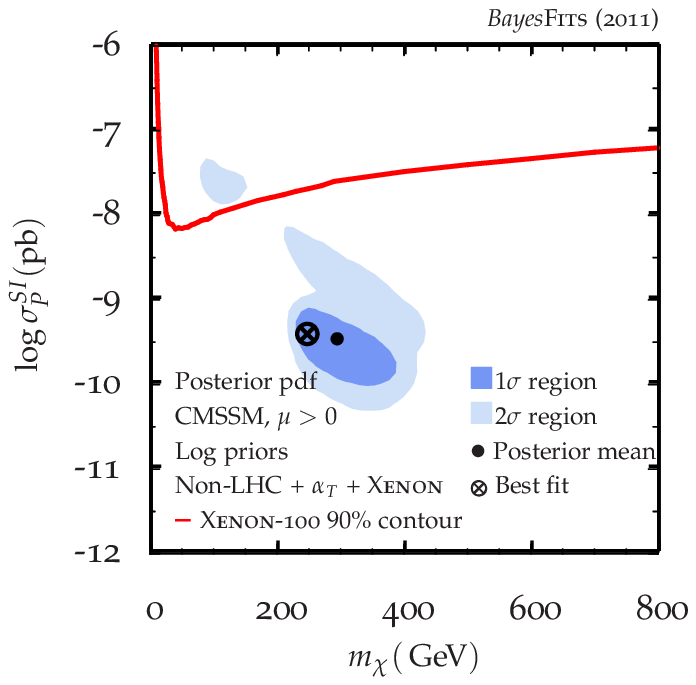}}
\caption[]{Marginalised posterior pdf on the (\sigsip,
  $m_\neutralinoone$) plane, \subref{fig:Comparison_sigsipmchi-a}
  before  and \subref{fig:Comparison_sigsipmchi-b} after we included a
  likelihood for the \xenon\ limit.}
\label{fig:Comparison_sigsipmchi}
\end{figure}

This can be seen in \reffigure{fig:Comparison_sigsipmchi} where we
present the impact of the \xenon\ 90\%~C.L. limit (denoted with 
solid red curve) on probability maps of the ($m_\neutone$, \sigsip) plane
for our scans. In \reffigure{fig:Comparison_sigsipmchi-a} the \xenon\
limit is not added to the likelihood, while in
\reffigure{fig:Comparison_sigsipmchi-b} it is applied. By comparing
both panels we can see that, once LHC limits have been applied (left
panel), the $1\sigma$ posterior region on the ($m_\neutone$, \sigsip)
plane is only weakly affected by the additional \xenon\ limit (right
panel).

Some interesting effects can nevertheless be noticed. First, a small
$2\sigma$ region above the \xenon\ exclusion curve has shrunk
somewhat, especially on the side of larger \sigsip, but, because of
the large theoretical error assumed in this analysis, it remains allowed. Second,
the large smearing affects the favored regions of \sigsip\ also below
the experimental curve.  Note that, before the \xenon\ is applied
(\reffigure{fig:Comparison_sigsipmchi-a}) there are actually two
$1\sigma$ regions close to each other. The lower one comes
entirely from the stau-coannhilation region of small \mzero\ and large
\mhalf. The other one, just above it (along
with a broader $2\sigma$ region, both decreasing with
$m_\neutralinoone$), corresponds to the broad $2\sigma$ AF region in
the (\mzero, \mhalf) plane. Once the \xenon\ limit is added to the
likelihood (\reffigure{fig:Comparison_sigsipmchi-b}), the largest
values of \sigsip, just below the experimental curve, become excluded
and the statistical significance of the whole AF region becomes
reduced to the $2\sigma$ level. On the other hand, the lower $1\sigma$
region remains nearly intact.

Clearly, recent LHC limits have had the effect of pushing down the
most favored ranges of \sigsip, for the most part below $\sim
10^{-9}\pb$, while pre-LHC data favored largest $1\sigma$ posterior
ranges of \sigsip\ at least an order of magnitude higher; compare,
\eg,\cite{Roszkowski:2007fd,Trotta:2008bp}. This implies much poorer
prospects for \xenon, with expected reach of $\sim
10^{-9}\pb$ to explore the SI cross sections currently favored by the CMSSM.

It is also clear that, in light of the impact of LHC limits, one-ton
detectors with planned sensitivity reach of $\sim10^{-10}\pb$ will now
be needed to explore the most probable ranges of the ($m_\neutone$,
\sigsip) plane in the CMSSM. We also note that the most probable range
of dark matter particle mass, $250\GeV\lsim m_\neutone \lsim 343\GeV$
(at $1\sigma$) is somewhat above the range of the highest sensitivity of
most detectors.  Finally, in \reftable{tab:mass_cred_regions}, we show
the combined impact of the \alphaT\ and xenon\ limits on the posterior
ranges of several particle masses already constrained by other
(Non-LHC) data.
The $1\sigma$ and $2\sigma$ posterior ranges were calculated with \refequation{eqn:cred_region_1D}.

\begin{figure}
\centering
\subfloat[][
]{%
\label{fig:NonLHC+alphaT+XENON_scan-a}%
\includegraphics[scale=\scale]{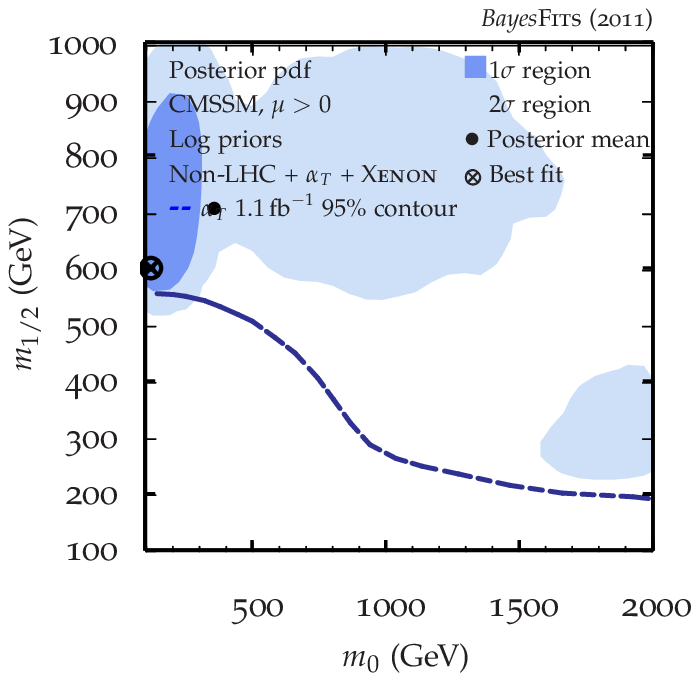}}%
\hspace{8pt}%
\subfloat[][
]{%
\label{fig:NonLHC+alphaT+XENON_scan-b}%
\includegraphics[scale=\scale]{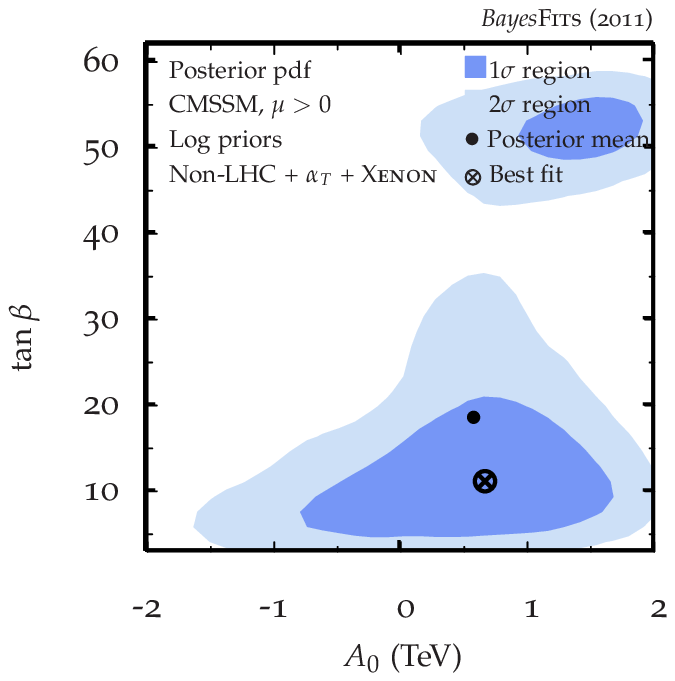}}%
\caption[]{Marginalised posterior pdf with the log prior the CMSSM's parameters constrained by non-LHC experiments,
  the \alphaTlim\ and the \xenon\ limit. The panels should be compared with the corresponding panels
in \protect\reffigure{fig:Comparison_m0m12_lhc} and \protect\reffigure{fig:Comparison_a0tanb_lhc}.}%
\label{fig:NonLHC+alphaT+XENON_scan}%
\end{figure}

Finally, in \reffigure{fig:NonLHC+alphaT+XENON_scan}, we show the
impact of the \xenon\ limit on the shapes of the 2D marginalized
posterior pdf maps
for the CMSSM parameters. The results shown here correspond to a scan
with a likelihood from the non-LHC constraints, from the \alphaTexp\
and from \xenon. The panels should be compared with the corresponding
panels in \reffigure{fig:Comparison_m0m12_lhc} and
\reffigure{fig:Comparison_a0tanb_lhc}.

The discussion of \reffigure{fig:Comparison_sigsipmchi} above helps in
understanding new effects in
\reffigure{fig:NonLHC+alphaT+XENON_scan}. First, it is clear that, 
the configurations of the CMSSM parameters are already constrained by current LHC limits to
give \sigsip\ for the most part below the \xenon\
limit. Unsurprisingly, adding it to the likelihood has a
relatively weak additional impact on the CMSSM's parameters.
The
$1\sigma$ posterior region in the high probability SC
region on the (\mzero, \mhalf) plane in
\reffigure{fig:NonLHC+alphaT+XENON_scan-a} remains unaffected because
the corresponding values of \sigsip\ are the lowest. The
change in the best-fit point is also probably not statistically
significant.
However, note that the broader $2\sigma$ region of AF has shrunk somewhat
on the side of large \mzero, especially in the direction of the HB/FP
region. This is because in that direction the Higgsino component of
the DM neutralino increases, causing in turn \sigsip\ to increase and
become at some point constrained by the \xenon\ limit, especially for
large \tanb.

Moving farther toward the lower-right corner, we can see 
that the \xenon\ result disfavors also the HB/FP region -- the $1\sigma$ credible region in the
(\mzero, \mhalf) plane in \reffigure{fig:Comparison_m0m12_lhc-a}
has now
become only a $2\sigma$ region. This is not surprising since predicted
values of \sigsip\ in that regions are among the largest predicted in
the CMSSM, 
again because of the increased Higgsino component of the neutralino.
We stress, however, that the HB/FP  region is not excluded by the
\xenon\ limit; rather, it is inside the $2\sigma$ credible
region. This follows from our conservative estimation of the error in
the \sigsip\ calculation. (We weakened the \xenon\ limit by smearing
it with a Gaussian describing the theoretical error in the \sigsip\
calculation.) We have checked that, if we assumed the theoretical
error in \sigsip\ to be $0.1\times \sigsip$, the \xenon\ limit would
exclude the focus point region at $2\sigma$.  Finally we note that the
favored regions on the (\azero, \tanb) plane in
\reffigure{fig:NonLHC+alphaT+XENON_scan-b} are not strongly affected
by \xenon, though the two modes on the (\azero, \tanb) plane are no
longer connected.

\begin{table}
\begin{tabular}{|c||c|c||c|c|}
\hline 
Mass (GeV) & $68$\%  & $95$\% & $68$\% & $95$\% \\
\hline 
& \multicolumn{2}{|c||}{Non-LHC}    & \multicolumn{2}{|c|}{Non-LHC +
  \alphaTlim\ + \xenon} \\
\hline 
$m_h$ & $(112.3,116.5)$ & $(110.1,118.4)$ & $(114.4,117.8)$ & $(112.2,119.4)$  \\
$m_{\neutralinoone}$ &$(56,291)$ & $(53,356)$ &$(250,343)$ & $(128,390)$\\
$m_{\charginoone}$ & $(110,554)$ & $(104,676)$ & $(475,651)$ & $(181,738)$\\
$m_{\squark}$ & $(326,808)$ &$(254,1172)$ & $(434,761)$ &$(398,1302)$\\
$\mgluino$ & $(403,1576)$ & $(384,1885)$ & $(1380,1825)$ & $(879,2043)$\\
\hline 
\end{tabular}\caption{Posterior $1\sigma$ and $2\sigma$ 
regions for several particle masses, when constrained by pre-
(two left columns) and post-$\alphaTonefb$ data,
calculated with
  \refequation{eqn:cred_region_1D}.  } 
\label{tab:mass_cred_regions}
\end{table}

\subsection{\label{sec:bfpoint}Prior dependence and the best-fit point}
We will now comment on the prior dependence of the  results presented
here, and will also come back to the discussion of the best-fit point.  

As is well known, some of the most challenging aspects of Bayesian
statistics are the necessity to choose a prior and the sensitivity of
the posterior to that choice.  The prior dependence of the posterior
is a measure of the lack of the constraining power of the likelihood
function. If the information from  data included in the likelihood is
sufficient to select ranges of model's parameters giving good fit to
the constraints (high posterior probability regions), then the
sensitivity of such ranges to the choice of priors should be weak, or
even  marginal. In such cases the likelihood has the ability
to overpower noninformative priors, such as log priors and linear
(flat) priors, and the resulting posterior has little prior
dependence. If, however, the posterior is dependent on the choice of
prior, the likelihood ought to be regarded as too weak to support
robust conclusions about the model.

In several pre-LHC studies prior dependence of the CMSSM was found to
be substantial (see, \eg,\cite{Roszkowski:2007fd,Trotta:2008bp}), and
the situation in less constrained models was found to be even less
satisfactory\cite{Roszkowski:2009sm}. This was merely a reflection of
poor constraining power of the data available at that time. It was
also concluded that the choice of the prior that is linear in the log
of the masses is more motivated by both physical and statistical
reasons\cite{Roszkowski:2007fd,Trotta:2008bp}. From the physical point
of view, log priors explore in much greater detail the low-mass
region, where one typically needs less fine tuning in order to achieve
radiative electroweak symmetry breaking. From the statistical point of
view, log priors give the same \textit{a priori} weight to all orders of
magnitude in the masses, and thus appear to be less biased to giving
larger statistical \textit{a priori} weights to the large mass region, which
under a flat prior has a much larger volume in parameter space.

\begin{figure}
\centering
\subfloat[][
]{%
\label{fig:NonLHC+alphaT+XENON_scan_flat-a}%
\includegraphics[scale=\scale]{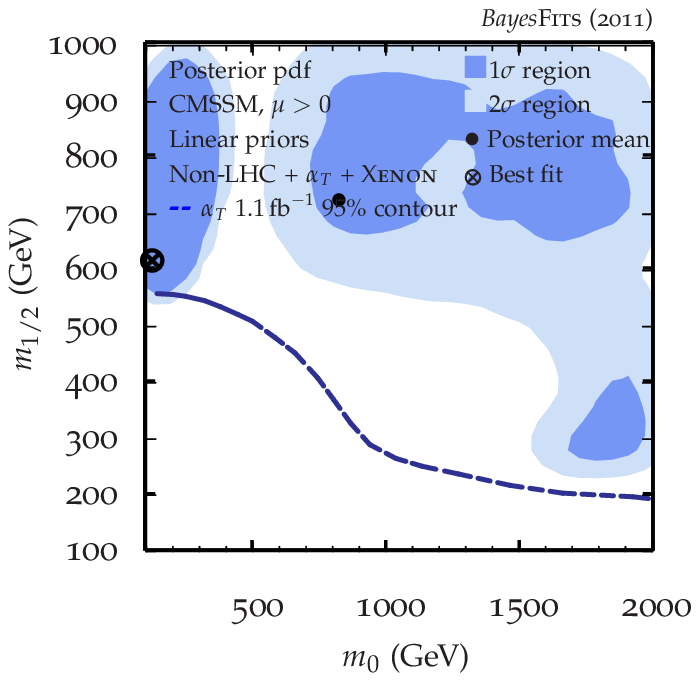}}%
\hspace{8pt}%
\subfloat[][
]{%
\label{fig:NonLHC+alphaT+XENON_scan_flat-b}%
\includegraphics[scale=\scale]{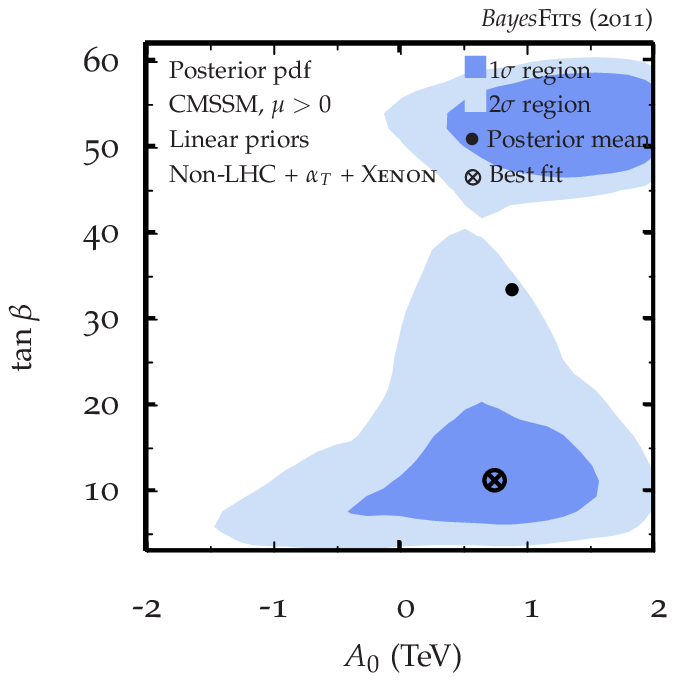}}%
\caption[]{Marginalised posterior pdf with the linear (flat) prior the CMSSM's
  parameters constrained by non-LHC experiments, the \alphaTlim\ and
  the \xenon\ limit. The Figure should be compared with
  \protect\reffigure{fig:NonLHC+alphaT+XENON_scan}.}%
\label{fig:NonLHC+alphaT+XENON_scan_flat}%
\end{figure}

Motivated by the above arguments, in this study we have chosen log
priors for the CMSSM's mass parameters, and our posterior
distributions are, of course, dependent on the choice. As a way of
examining the degree of the dependence, we identically repeated our
scan of the CMSSM with a likelihood from the non-LHC experiments and
from \alphaT, except that this time we chose linear priors for all the
CMSSM's parameters. 
The resulting posterior distributions, which are
shown in \reffigure{fig:NonLHC+alphaT+XENON_scan_flat}, are 
actually broadly similar to the distributions that resulted from log
priors. 
With linear priors, we found
three $1\sigma$ modes in the marginalized posterior pdf on
the (\mzero, \mhalf) plane, rather than two as in
\reffigure{fig:Comparison_m0m12_lhc-b}. The third mode is present in
\reffigure{fig:NonLHC+alphaT+XENON_scan_flat} both large \mhalf\ and
\mzero, and  probably results from the mentioned above volume effect
on the marginalized posterior. For the same reason the $2\sigma$ region at
large masses also becomes enhanced and the FP/HB mode again becomes
strengthened.
We interpret this as a reflection of the fact that, despite new strong
limits from the LHC, the constraining power of experimental data
remains insufficient to eliminate prior dependence of the CMSSM. On
the other hand, the effect of assuming the linear prior for \mhalf\ and
\mzero on the parameters \tanb\ and \azero\ appears to be relatively
weaker, as expected, since for the latter the prior has been assumed
to be the same (linear) in both cases.

\begin{table}
\begin{tabular}{|c|c|c|c|c|c|c|c|c|c|c|c|c|c|c|c|c|c|}
\hline 
Constraints 			& \mzero & \mhalf & \azero & \tanb & $\chi^2$ & d.o.f. & $p$-value\\
\hline 
Non-LHC 			& $122$ $(116, 1391)$ & $343$ $(142,
702)$ & $806$ $(236, 1514)$ & $17$ $(13, 22)$ & $16.53$ & $16$ &
$42\%$\\ 
\hline
%
Non-LHC + \alphaT\ + \xenon 	& $122$ $(127, 741)$ & $600$
$(608, 820)$ & $677$ $(82, 1283)$ & $11$ $(9, 16)$ & $22.21$ & see text & see text \\
\hline 
\end{tabular}
%
%
%
%
\caption{Best-fit points and $68\%$ central credible regions for the CMSSM's parameters,  calculated with
  \refequation{eqn:cred_region_1D}, when constrained by different sets of experimental data. Masses are in \GeV. 
  The best-fit values of \mzero\ are close to the lower edge of our
  prior range for \mzero, which is $100\GeV$; therefore,  
  our central credible regions exclude the best-fit values of \mzero.}
\label{tab:bestfit_params}
\end{table}

\begin{table}
\footnotesize{
    \begin{tabular}{|c|c|c|c|c|c|c|c|c|c|c|c|c|c|c|}
\hline 
Constraint & \abundchi & \mh & \bsg & \sineff & \mw & \deltaa &  \btn & \deltam & \alphaT &  Total\\ \hline 
$\chi^2$ & $0.01$ & $1.32$ &  $1.98$ & $4.16$ & $1.49$ & $9.76$ & $0.03$ & $0.25$ & $3.42$ & $22.21$\\
\hline 
\end{tabular}
}
%
%
%
\caption{Breakdown of the main contributions to $\chi^2$ for our
  best-fit point from a scan with a likelihood from  non-LHC
  experiments and the \alphaT\ and \xenon\ limits. Note that all likelihoods are
  normalized to unity, including one for each $H_T$
  bin. }
\label{tab:bestfit_lnlike}
\end{table}

On the other hand, the location of the best-fit point should, by
definition, be entirely determined by the likelihood function, and
therefore independent of the prior. However, because in practice it is
found numerically with a Monte Carlo method, it is a random variable
with an error and with a weak dependence on the scanning algorithm,
rather than an exact solution. This is clear when one remembers that
by choosing, for example, the log prior for the CMSSM's mass
parameters, one in practice also chooses a log metric\footnote{We say
  metric, rather than prior, to stress that this effect is not related
  to our choice of Bayesian statistics.} for these parameters. As a
result, the scanning algorithm inevitably explores the low-mass region
of the CMSSM's (\mzero, \mhalf) plane in greater detail than the
high-mass region. This metric dependence can be overcome, or at least
reduced, by tightening the algorithm's stopping conditions, so that it
runs until it has explored the whole parameter space in sufficient
detail. We also emphasise once again that, in Bayesian statistics the
best-fit point has no significance. 

In \reftable{tab:bestfit_params} we present best-fit points from two
scans of the CMSSM: one with non-LHC constraints only and one with  \alphaT\ and \xenon\
constraints added to the likelihood. (We quote our best-fit points rounded
to the nearest whole \GeV\ or unit of \tanb.) In both cases the
best-fit point is located in the SC/AF region of the CMSSM's parameter
space. Our best-fit point for the non-LHC-only case is in good agreement with
best-fit points reported previously in \refref{deAustri:2006pe,Allanach:2007qk,Trotta:2008bp,Buchmueller:2010ai}, which
is encouraging. Our $p$-value of $42\%$ is also reasonably close to
MasterCode's latest result for the CMSSM of $37\%$ for this
case\refref{Buchmueller:2011sw}.

On the other hand, after including the \alphaT\ and \xenon\
constraints in the likelihood, our best-fit point shifts up
almost vertically to a larger value of $\mhalf\simeq600\GeV$, while
the one in \refref{Buchmueller:2011sw} moves much more radically, to
much larger values of both mass parameters ($\mzero=450\GeV$ and
$\mhalf=780\GeV$), and also giving much larger $\tanb=41$, although
with large standard deviations reported for all the parameters.

We investigated which constraints included in the likelihood played the most
important role in determining the shape of the posterior, and
also the location of the best-fit point.  In
\reftable{tab:bestfit_lnlike} we show a breakdown of the main
contributions to the $\chi^2$ for
the best-fit point corresponding to including the \alphaT\ and \xenon\
constraints in the likelihood. (Note that the \alphaT\ likelihood is
itself a product of eight likelihoods, and therefore contributes 8 degrees
of freedom, and inevitably has a relatively poor $\chi^2$.) It is
clear that the constraint from \deltaa\ plays the biggest role in
increasing the $\chi^2$ for
the best-fit point.

\begin{figure}
\centering
\subfloat[][\deltaa]{%
\label{fig:3d_bsg_g2-a}%
\includegraphics[scale=\scale]{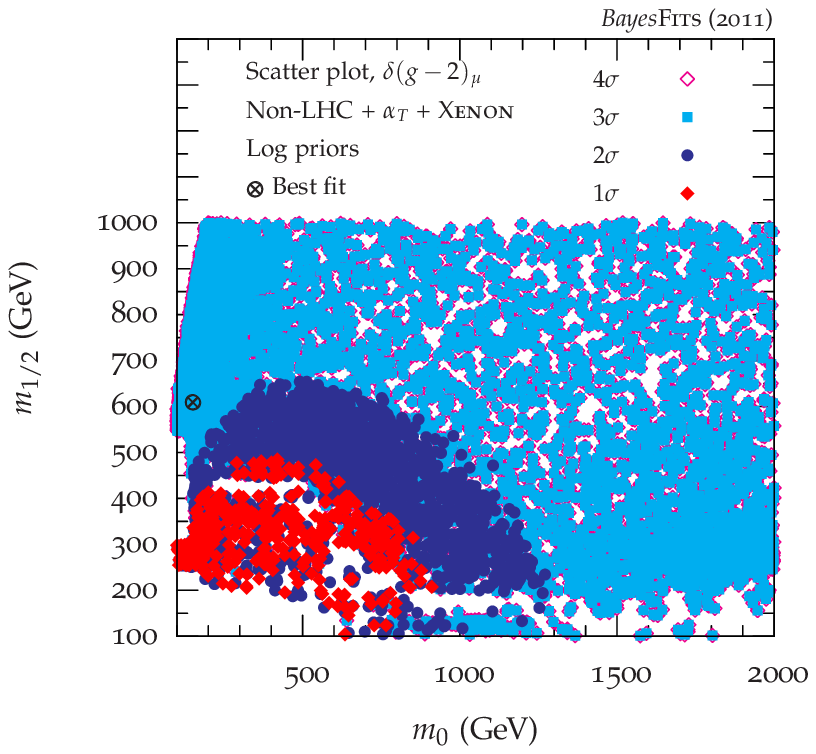}}%
\hspace{8pt}%
\subfloat[][\deltaa]{%
\label{fig:3d_bsg_g2-b}%
\includegraphics[scale=\scale]{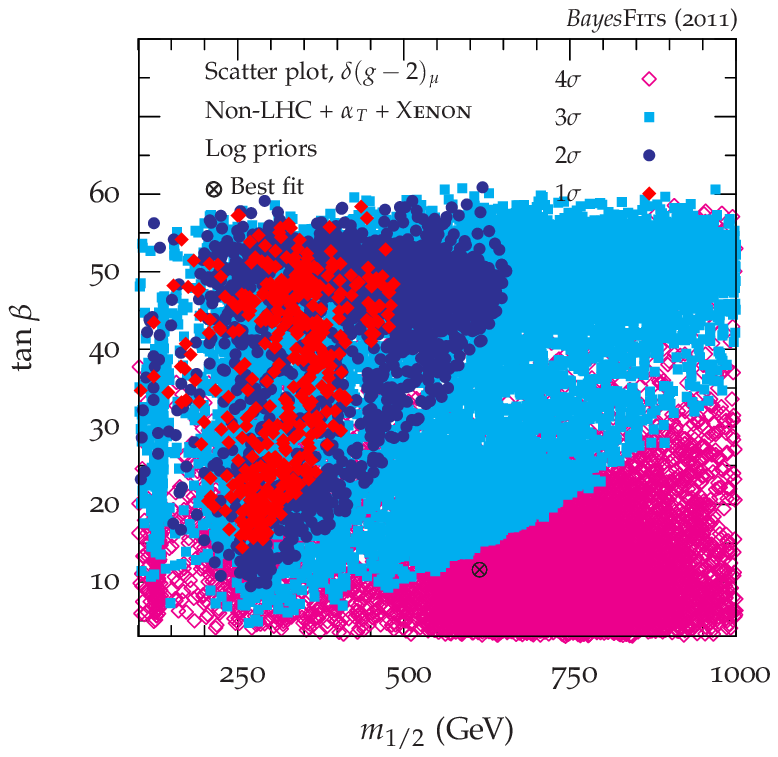}}\\
\subfloat[][\brbsg]{%
\label{fig:3d_bsg_g2-c}%
\includegraphics[scale=\scale]{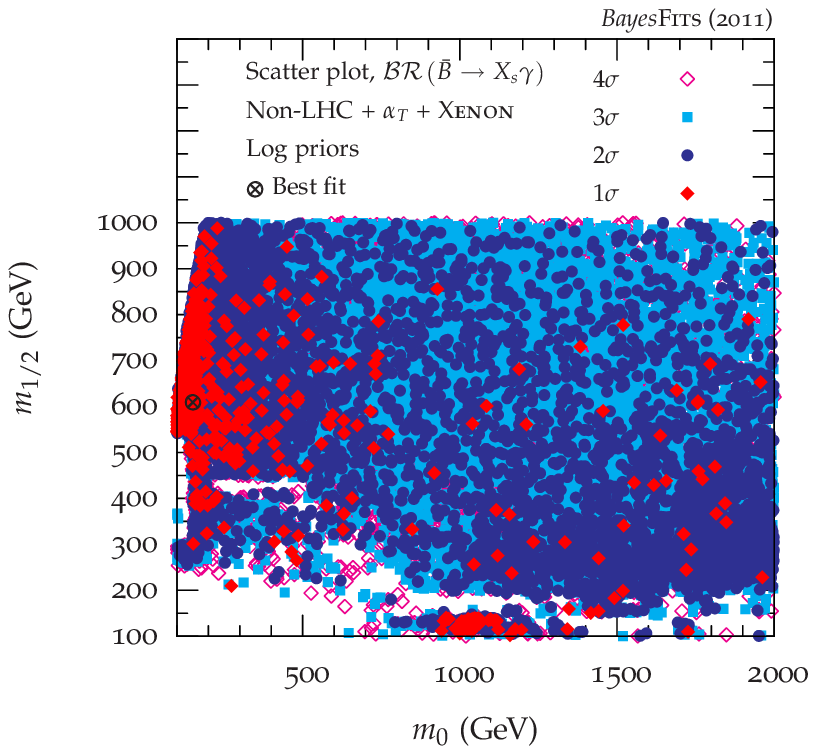}}%
\hspace{8pt}%
\subfloat[][\brbsg]{%
\label{fig:3d_bsg_g2-d}%
\includegraphics[scale=\scale]{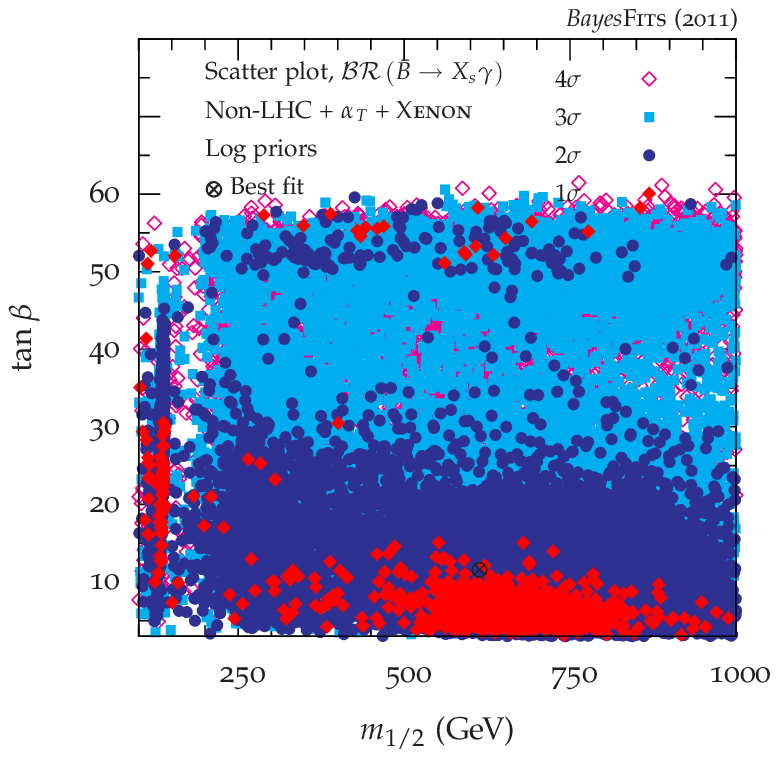}}
\caption[]{Scatter plots of points from a scan with a likelihood from the
  \alphaT\ and \xenon\ limits and from non-LHC experiments, colored by the values of
  \deltaa\ (top row) and \brbsg\ (bottom row). The colors show the
  discrepancy between the CMSSM's predicted value and the central experimental
  value, measured in experimental errors.}%
\label{fig:3d_bsg_g2}%
\end{figure}

The upper panels of \reffigure{fig:3d_bsg_g2} show that the \deltaa\
constraint requires that increases in \mhalf\ induced by the \alphaT\
limit are compensated by increases in \mzero\
[\reffigure{fig:3d_bsg_g2-a}], and also in \tanb\
[\reffigure{fig:3d_bsg_g2-b}]. Our best-fit point is, however, being
pulled in a different direction mostly by \brbsg. The lower panels of
\reffigure{fig:3d_bsg_g2} show that increasing \mhalf\ by improving
LHC mass limits, does not require one to increase \mzero\
[\reffigure{fig:3d_bsg_g2-c}] and small \tanb\ is sufficient
[\reffigure{fig:3d_bsg_g2-d}], in order to maintain a good fit to
\brbsg. There is a clear tension between on the one hand \deltaa,
which favors lighter mass spectra in order to generate large enough
SUSY contribution to the variable, and, on the other hand, \brbsg\ and
other constraints which prefer a SUSY contribution to be all but
suppressed.\footnote{The tension between \deltaa\ and \brbsg\ has
  already been investigated in \refref{Roszkowski:2007fd} for the
  pre-LHC case.} The tension between the two observables is
exacerbated by adding the constraint from \alphaT, because it pushes
\mhalf\ into a region of the (\mzero, \mhalf) plane in which it is
more difficult to satisfy both constraints simultaneously. It appears that, at the
end \deltaa\ is outweighed by the other constraints, most notably
\brbsg. As a result, we find that the best-fit point remains at small
\mzero\ and \tanb, and \deltaa\ is forsaken. Consequently, the
best-fit points have a large $\chi^2$ from \deltaa but not from
\brbsg; compare \reftable{tab:bestfit_lnlike}.

To investigate its effect, we repeated our analysis without the
\deltaa\ experimental constraint. With a likelihood from non-LHC
experiments, except for \deltaa, and from \alphaT, the credible
regions on the (\mzero,\mhalf) plane  are similar to those in
\reffigure{fig:Comparison_m0m12_lhc-b}, though the central region of
the plane is disfavored. The best-fit point's value of \mhalf\ is
significantly larger than that in
\reffigure{fig:Comparison_m0m12_lhc-b}.  This suggests that, while it
favors smaller values of the SUSY mass parameters, \deltaa\ 
does not play the dominant role 
in determining the credible regions. 

\begin{figure}
\centering
\subfloat[t][\brbsg\ against \deltaa.]{%
\label{fig:2d_bsg_g2_bsmm_oh2-a}%
\raisebox{1mm}{\includegraphics[scale=\scale]{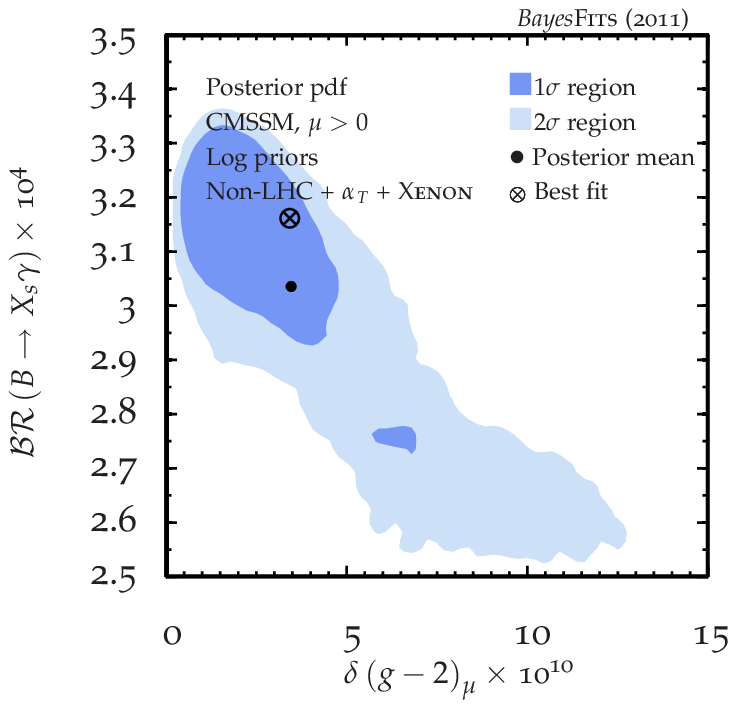}}}%
\hspace{8pt}%
\subfloat[t][\brbsmumu\ against \brbsg.]{%
\label{fig:2d_bsg_g2_bsmm_oh2-b}%
\raisebox{3.5mm}{\includegraphics[scale=\scale]{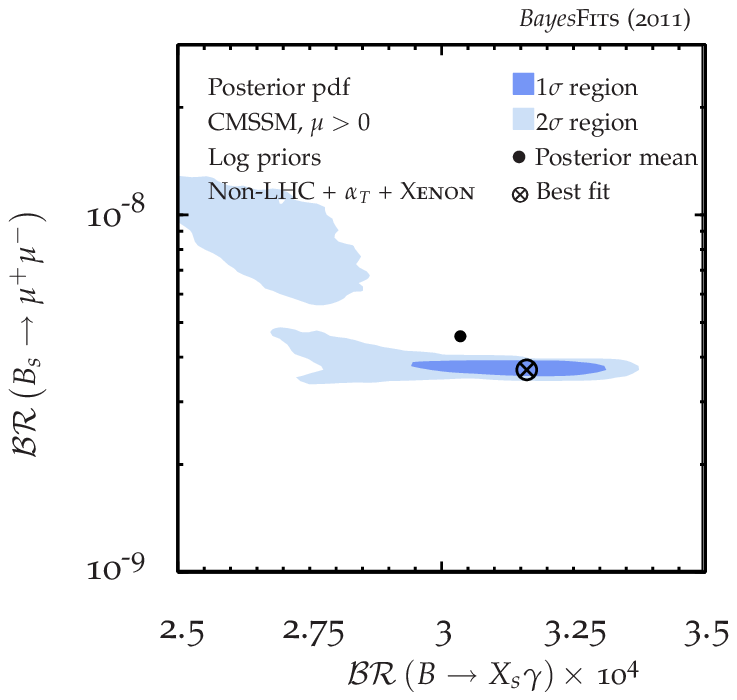}}}
\hspace{8pt}%
\subfloat[t][\brbsmumu\ against \deltaa.]{%
\label{fig:2d_bsg_g2_bsmm_oh2-c}%
\raisebox{1.5mm}{\includegraphics[scale=\scale]{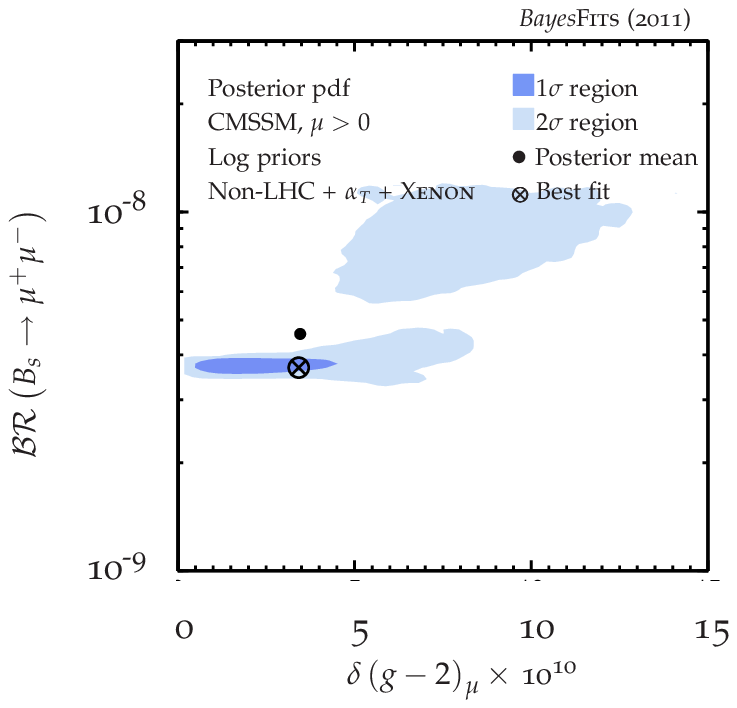}}}
\hspace{8pt}%
\subfloat[t][\brbsg\ against \relic.]{%
\label{fig:2d_bsg_g2_bsmm_oh2-d}%
\includegraphics[scale=\scale]{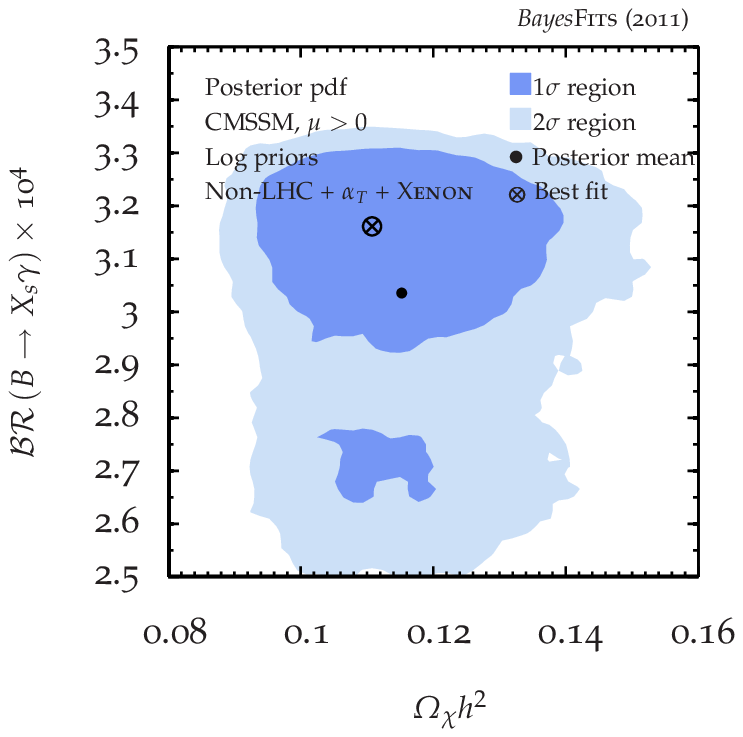}}\\%
\caption[]{Marginalised posterior pdf for combinations of the experimental observables \brbsg, \deltaa, \brbsmumu\ and \relic, from a scan with a likelihood from \alphaT,  \xenon\ and non-LHC experiments.}%
\label{fig:2d_bsg_g2_bsmm_oh2}%
\end{figure}

This is further illustrated in of
\reffigure{fig:2d_bsg_g2_bsmm_oh2-a}, where we show the interplay
between combinations of constraints from \brbsg, \brbsmumu, \deltaa\
and \abundchi\ in 2D marginalized posteriors. We can also see that the
resulting range of \brbsmumu\ is very close to its SM value and that
\abundchi\ plays a rather neutral role in determining both the
posterior and the best-fit point. This is because one can relatively
easily adjust \azero\ to produce the correct value of \abundchi,
without much affecting the other major constraints.

Another note is in order about $\chi^2$ and the $p$-value. As can be seen from
\reftable{tab:bestfit_params}, applying only non-LHC constraints, we
find $\chi^2=16.53$  and the $p$-value of $42\%$.  Adding new constraints from the \alphaT\ and
\xenon\ likelihood inevitably increases the value of $\chi^2$, but a shift in
the $p$-value depends on our assumptions about the additional degrees
of freedom.  In the CMS analysis\cite{CMS:2011} the observed events in
each bin are treated as independent of the other bins, and described by a Poisson
distribution. This is why in our treatment of the \alphaT\
limit we evaluated likelihood in each of the eight energy bins
independently, treating them as contributing as many new degrees of freedom. In this case the $p$-value
in fact increases to $61\%$. On the other hand, one could
reasonably expect that signal events in different bins would be
correlated, in which case the \alphaT\ constraint could be treated as effectively
contributing only one extra degree of freedom. In this case the resulting
$p$-value would be 22.3\%. This illustrates the difficulty in
correctly estimating the number of degrees of freedom and the
$p$-value.  We also note that in a recent analysis of the
MasterCode\cite{Buchmueller:2011sw} the $p$-value diminished from
$37\%$ to $15\%$ when the LHC and \xenon\ constraints were
included. Note, however, that in that analysis Gaussian distributions
were assumed, unlike here. Also, their way of including LHC limits from
jets+\ETslash in the likelihood contributed only 1 additional degree
of freedom and also forward-backward scattering variables
$A_\text{fb}$ were included
in their likelihood, which suppressed their $p$-values further. All
this makes it difficult to compare the resulting values of $\chi^2$
and $p$-value.\\


In addition to a different location of the best-fit point and the
$p$-value, the confidence intervals on the
(\mzero, \mhalf) plane that are found in \refref{Buchmueller:2011sw} are much larger than our confidence
intervals and  span
the central region. On the other hand, they do not
include the focus point region. In contrast, we find regions of the
largest posterior not only in the SC/AF region but also in the focus
point region, both in the pre-LHC case, in agreement with previous
studies using the SuperBayeS code\cite{Trotta:2008bp}, and also after
including the \alphaT\ and \xenon\ limits. A physical reason for its
existence has been given earlier.

The apparent discrepancy between some of the results reported here and
those in a recent $\chi^2$ analysis by the MasterCode
group\cite{Buchmueller:2011ki} is probably caused by a combination of
several factors. Some have already been mentioned above. Most likely,
the different ways of implementing LHC limits play an important
role.\footnote{L.R.~and J.~Ellis.~(private communication)}  We
also note that the implementation of \brbsg\ in the version of
SuperBayeS that we have used for this analysis differs from that of
the MasterCode group whose experimental constraint is the ratio of the
measured branching ratio to its Standard Model prediction. In
contrast, in SuperBayeS the constraint from \brbsg\ is simply applied
to the measured branching ratio. On the other hand, the choice of
different statistics (Bayesian vs $\chi^2$) is probably of secondary
importance since our results for the best-fit point agree reasonably
well in the non-LHC case with several other analyzes, both Bayesian
and
$\chi^2$,\cite{Trotta:2008bp,Allanach:2011ut,Bechtle:2011it,Buchmueller:2011aa}.
We also note that the large $1\sigma$ errors reported on the CMSSM
parameters for the best-fit point in \cite{Buchmueller:2011ki}
indicate to us that probably small differences in the likelihood
function may lead to large shifts in the location of the best-fit
point, in addition to numerical issues related to using different
scanning algorithms. This may imply that the high probability regions
of the CMSSM parameter space after including current LHC constraints
are quite ``unstable,'' or can fairly easily shift within a rather
wide plateau, in the sense that ``secondary'' issues such as the
precise implementation of some of the main constraints, or a treatment
of LHC limits may lead to major differences, and seemingly
contradictory results.

\section{\label{sec:summary}Summary}
We have performed an updated Bayesian analysis of the CMSSM. In
addition to updating experimental inputs in indirect modes of
constraining SUSY [most notably \brbsmumu], we included much improved
limits from the \alphaTexp, which currently gives the strongest bounds
on the CMSSM mass parameters, and from the \xenon\ experiment. We
simulated the \alphaTexp\ in an approximate but methodologically
correct way by estimating the efficiency for the \alphaT\ method and
constructing a likelihood function. We validated our method against
the official \cms\ $95\%$ contour. For the \xenon\ limit we
constructed a conservative, approximate likelihood function by taking
into account large uncertainties related to the inputs to hadronic
matrix elements and the local density of dark matter.

We incorporated these likelihoods into a global Bayesian fit of the
CMSSM and identified marginalized posterior maps of the CMSSM's
parameter space. These credible regions were compared with credible
regions before the new experiments to illustrate the effects of the
new constraints. The \alphaT\ limit has taken a deep bite into regions
of the CMSSM's parameter space that were previously favored, and has
pushed the best-fit point to significantly higher values of
\mhalf. Including \xenon\ in the likelihood had a weak additional
effect.  We find that, although the focus point region is disfavored
by the new constraints, it is still not excluded.

Despite the disappointing null results of SUSY searches at the LHC so
far, which have pushed the allowed ranges of CMSSM mass parameters,
especially \mhalf, up to much larger values, we note that the best-fit
point, both before and after including the \alphaTlim, is found close
to the bottom of the marginalized $1\sigma$ posterior ranges of both
\mzero\ and \mhalf. Despite the uncertainties in determining its
location discussed above, this is certainly encouraging for prospects
of finding a signal of SUSY in a much larger dataset already collected
by both \atlas\ and \cms. On the other hand, we note that the
preference for the low-mass spectrum in the CMSSM and similar unified
models is driven primarily by a single constraint from the \deltaa\
anomaly, satisfying what has already been in some tension with \brbsg, and is
now becoming increasingly harder also with improving LHC limits.

\begin{acknowledgments}
  A.J.F.~is funded by the Science Technology and Facilities
  Council. L.R. and S.T. are funded in part by the Welcome Programme
  of the Foundation for Polish Science. L.R. is also supported in part
  by the Polish National Science Centre Grant No.~N202 167440, an STFC
  consortium grant of Lancaster, Manchester and Sheffield Universities,
  and by the EC 6th Framework Program No.~MRTN-CT-2006-035505.
\end{acknowledgments}

\bibliography{bf1}

\end{document}